%% file: main.tex
\documentclass[preprint]{iacrtrans}

\usepackage{epsfig}
\usepackage{amssymb}
\usepackage{amsmath}
\usepackage{multicol}
\usepackage{colortbl}
\usepackage{booktabs}
\usepackage{multirow}
\usepackage[subtle]{savetrees}
\usepackage[utf8]{inputenc}
\usepackage{amsfonts}
\usepackage{graphicx}
\graphicspath{ {images/} }
\usepackage{algorithmic}
\usepackage[boxruled,vlined,linesnumbered]{algorithm2e}
\usepackage[n ,advantage ,operators ,sets, adversary,landau,probability,notions,logic,ff,mm,primitives,events,complexity,asymptotics,keys]{cryptocode}
\usepackage{caption}
\usepackage{float}
\usetikzlibrary{matrix,shapes,arrows,positioning,chains, calc}
\usepackage{tikz}
\usetikzlibrary{fit}
\usetikzlibrary{shapes,arrows,positioning}
\newlength{\nodedistance}
\usepackage{amsmath}
\usepackage{amsthm} 
\usepackage{hyperref}
\usepackage{centernot}
\usepackage{lscape}
\usepackage{enumitem}
\usepackage{bigdelim}
\usepackage{adjustbox}
\usepackage{subcaption}
\usepackage{bm}

\SetKwData{Left}{left}\SetKwData{This}{this}\SetKwData{Up}{up}
\SetKwFunction{Union}{Union}\SetKwFunction{FindCompress}{FindCompress}
\SetKwInOut{Input}{Input}\SetKwInOut{Output}{Output}

\def\x{{\mathbf x}}
\def\y{{\mathbf y}}

\newcommand{\seeda}{\ensuremath{\text{seed}_{\pmb{A}}}}
 
\newcommand{\KEM}{\ensuremath{\mathtt{KEM}} }

\newif\ifsubmission
\submissionfalse

\ifsubmission

\else
\authorrunning{Kundu et al.}
\fi

\title[Carry Your Fault]{Carry Your Fault: A Fault Propagation Attack on Side-Channel Protected LWE-based KEM}

\begin{document}

\author{Suparna Kundu\inst{1}, Siddhartha Chowdhury\inst{2}, Sayandeep Saha\inst{3}, Angshuman Karmakar\inst{1,4}, Debdeep Mukhopadhyay\inst{2} \and Ingrid Verbauwhede\inst{1}}
\institute{
  COSIC, KU Leuven, Belgium
  \and
  Indian Institute of Technology Kharagpur, India
  \and
  Université catholique de Louvain, Belgium
  \and
  Indian Institute of Technology Kanpur, India\\
  \email{{suparna.kundu,angshuman.karmakar,ingrid.verbauwhede}@esat.kuleuven.be}
  \email{{siddhartha.chowdhury92,sayandeep.iitkgp,debdeep.mukhopadhyay}@gmail.com}
}

\maketitle


\begin{abstract}

Post-quantum cryptographic (PQC) algorithms, especially those based on the learning with errors (LWE) problem, have been subjected to several physical attacks in the recent past. Although the attacks broadly belong to two classes -- passive side-channel attacks and active fault attacks, the attack strategies vary significantly due to the inherent complexities of such algorithms. Exploring further attack surfaces is, therefore, an important step for eventually securing the deployment of these algorithms. Also, it is important to test the robustness of the already proposed countermeasures in this regard.

In this work, we propose a new fault attack on side-channel secure masked implementation of LWE-based key-encapsulation mechanisms (KEMs) exploiting fault propagation. The attack typically originates due to an algorithmic modification widely used to enable masking, namely the Arithmetic-to-Boolean ($\mathtt{A2B}$) conversion. 
We exploit the data dependency of the adder carry chain in $\mathtt{A2B}$ and extract sensitive information, albeit masking (of arbitrary order) being present. As a practical demonstration of the exploitability of this information leakage, we show key recovery attacks of Kyber, although the leakage also exists for other schemes like Saber. The attack on Kyber targets the decapsulation module and utilizes Belief Propagation (BP) for key recovery. To the best of our knowledge, it is the first attack exploiting an algorithmic component introduced to ease masking rather than only exploiting the randomness introduced by masking to obtain desired faults (as done by Delvaux). Finally, we performed both simulated and electromagnetic (EM) fault-based practical validation of the attack for an open-source first-order secure Kyber implementation running on an STM32 platform.

\end{abstract}


\input{sections/introduction}
\input{sections/preliminaries}
\input{sections/fault_attack}
\input{sections/relax_fa}

\input{sections/experiment}
\input{sections/discussions}

\input{sections/conclusion}

\bibliographystyle{alpha}
\bibliography{main}


\end{document}

%% file: sections/introduction.tex
\section{Introduction}\label{sec:intro}
Post-quantum cryptography (PQC) comprises cryptosystems that are designed using hard problems that an adversary cannot solve \emph{easily} using a sizeable quantum computer. PQC has been developed as a contingency plan in anticipation of the invention of sizeable quantum computers that can break our prevalent classical cryptosystems in the near future. A remarkable step in this direction is the recently concluded post-quantum standardization procedure initiated by the National Institute of Standards and Technology (NIST). NIST proposed some standard Digital Signatures, Key-Encapsulation Mechanisms (KEM), and Public-Key Encryption schemes~\cite{nist_final_report}, which can be used to replace our current cryptosystems.
Therefore, we should put more focus on practical issues such as the physical security of PQC so that it generates enough confidence to be deployed in the near future.

Countermeasures for two different types of physical attacks \textit{i.e.} side-channel attacks (SCA) and Fault Attacks (FA) differ greatly from each other. Usually, countermeasures for one type of physical attack do not guarantee any security against another type of physical attack. \textcolor{black}{However, if an SCA countermeasure is not applied carefully, its application sometimes opens up new vulnerable points for SCA. One such attack is shown in~\cite{FouqueRVD08}. This paper exploited carry leakage from the randomized exponent of RSA.}
Masking~\cite{ChaJutRaoRoh1999} is a well-known and provably secure countermeasure against SCA. 
Masking alone provides no assurance of security against FAs, except in some special cases of Statistical Ineffective Fault Attack (SIFA)~\cite{saha2019framework}\footnote{Masking alone provides some SIFA security only for cases where faults cannot be injected inside S-Boxes and the faulty outputs are blocked leaving SIFA as the only option (ruling out Differential Fault Attacks aka. DFAs). However, this guarantee does not hold when faults can corrupt the S-Box internals and fine-grained error correction is needed along with masking.}. However, protection against both attacks is desired and, therefore, it is important to understand the robustness of each dedicated countermeasure against the other attack vectors, \textit{e.g.} masking against FAs. It is especially important to understand whether one such countermeasure results in some new vulnerabilities that are exploitable by attack vectors. Recent work on symmetric key ciphers has shown that the interaction between two countermeasure classes is tricky and often leads to new attacks~\cite{saha2021divided,saha2023non}. We find that there is a lack of study on such \textit{cross-attack} strategies \textit{i.e.} FAs on schemes with SCA countermeasures or vice versa for PQC schemes. Therefore, in this work, we ask: \textit{Does one countermeasure create new vulnerabilities for another class of attacks in the PQC context?} Specifically, since there exist many SCA secure masking schemes \cite{FO-masked-saber,ho-mask-comparator-kyber-bos-2021,FO-masked-kyber-HeinzKLPSS22,HO_mask_Saber} for learning with errors (LWE)~\cite{DBLP:journals/jacm/Regev09}-based PQC key-encapsulation mechanisms (KEM), in this work, we will limit our focus to these schemes only.

Most of the LWE-based KEMs (e.g. Kyber~\cite{Kyber-Kem}, Saber~\cite{Saber_kem}, NewHope \cite{newhope}, etc.) use a variant of the Fujisaki-Okamoto (FO)~\cite{FujisakiO99} transformation~\cite{Jiang2017} to achieve security against Chosen Ciphertext Attack (CCA) attacks. A CCA-secure KEM consists of three algorithms: Key generation, Encapsulation, and Decapsulation. Although the key generation and encapsulation algorithms can be attacked using fault or side-channel~\cite{DBLP:conf/cosade/RaviRBCM19,DBLP:conf/hipeac/ValenciaOGR18}, the decapsulation algorithm is especially exploitable as it contains the secret key~\cite{DBLP:conf/latincrypt/PesslP19}. The secret key of KEM is non-ephemeral \textit{i.e.} the same secret key is repeatedly used in the decapsulation for different ciphertexts from various sources. Therefore, an attacker can refine their attack and improve their success probability by observing multiple traces with different inputs or by applying the fault in different executions while the key remains fixed. The decapsulation operation includes decryption and re-encryption and only returns a session key if the given ciphertext and re-encrypted ciphertext are the same. Otherwise, it returns a random key indicating decryption failure. These steps of the FO transformation inherently resist DFA~\cite{1st-order_masked_comparator, DBLP:conf/wistp/TunstallMA11}, which require faulty outcomes to proceed. However, a fault to bypass the equality check (between the received ciphertext and the re-encrypted ciphertext) can enable DFA. While this attack is simple, it is not the only attack possible. More precisely, equality check is not the only place where faults can be injected. Generally, ineffective FAs, similar to SIFA~\cite{saha2019framework} and Fault Template Attacks (FTA)~\cite{saha2020fault}, have been quite successful so far~\cite{PesslP21,HermelinkPP21,Delvaux22}. \textcolor{black}{Such attacks extract information from two events: a) in case there is decryption failure due to faults (effective fault), and b) the decryption succeeds even though a fault has been injected (ineffective fault). The ``effectiveness'' of a fault depends on the decryption noise, which depends on the long-term secret key $\pmb{s}$. Therefore, each decryption failure or success provides an inequality involving $\pmb{s}$. The attacker can accumulate a sufficient number of such inequalities by injecting several faults and can retrieve $\pmb{s}$ by solving this system of inequalities.}

Several works~\cite{DBLP:conf/fdtc/BettaleMR21,PesslP21,HermelinkPP21,Delvaux22} have proposed efficient ineffective FAs on LWE-based KEM schemes. In~\cite{PesslP21}, Pessl et al. used an instruction-skip fault in the $\mathtt{Decode}$ part (\autoref{fig:kyberfunc}) of the decapsulation of Kyber and then observed if this fault resulted in the correct session key or not. 
The attack proposed in~\cite{HermelinkPP21} is similar to~\cite{PesslP21} but uses a different location to inject fault, which also belongs to the category of ineffective fault attack. Later, Jeroen Delvaux extended the fault locations of~\cite{HermelinkPP21} and improved the inequality solver in~\cite{Delvaux22}. There is another recent class of attack (not ineffective faults), which injects faults in the constants of the Number-Theoretic-Transform (NTT) computation, reducing the entropy of the LWE instances leading to key recovery~\cite{RaviYBZC23}. Notably, in all of these cases, the main difference comes from the fault location -- finding out new attack location is always beneficial from a security assessment perspective. It also indicates where one should put the countermeasures and what kind of countermeasures are needed.    
\textcolor{black}{
In this regard, the contributions of this paper are as follows:
\begin{itemize}
\item We propose an FA by injecting faults in a previously unexplored location present in many LWE-based KEMs. The attack falls in the category of ineffective FAs. However, the new location we explore (the Arithmetic-to Boolean mask conversion module aka. $\mathtt{A2B}$) is an artifact of masking. Therefore, our attack can be seen as a ``side-effect'' of masking. {There exists a previous attack due to~\cite{Delvaux22} which is aided by masking. The reason was that masking randomizes the variable to be faulted and increases the chance of realizing a desired (e.g. single-bit) fault model with different executions. Our exploration is different -- we show that an algorithmic change, which is essential for efficient masking, is problematic. Also, we shall later in this paper that preventing these two attacks have very different requirements, and anticipate that our attack would be difficult to prevent using standard techniques, such as duplication (in Sec.~\ref{sec:discussion}).}
\item The proposed vulnerability originates from a fault-induced information-leakage channel through the carry propagation logic of a masked adder. While we restrict our discussion mostly to Kyber in this paper, the channel exists for Saber (we will explain this briefly) and maybe for many other schemes that utilize such adders. Moreover, the leakage is oblivious to the masking order, and, therefore, the attack remains unaffected by the increase in SCA security.
\item We realize the key recovery algorithm using the Belief Propagation algorithm provided in~\cite{Delvaux22}; {however, with important customizations needed for our purpose}. While the exact version of the attack requires a single-bit fault at a specific bit, we show that the effectiveness remains unchanged for much relaxed multi-bit fault models -- although at the cost of more observations. {This interesting observation is due to the nature of carry propagation through the adder and the restricted range and distribution of values the message coefficients can take.}  
\item Our final contribution is a practical realization of the attack on an STM32-based open-source first-order masked implementation of Kyber. We used electromagnetic (EM) faults for this purpose and effectively recovered the key with {$1.9$} million injections.  
\end{itemize}
{It is worth mentioning that there exist multiple works targeting the masked implementations of KEMs using SCA analysis~\cite{NgoDGJ21,NgoWDP22}. Most of them, such as~\cite{NgoDGJ21}, exploit the inherent weaknesses of masked software implementations of these algorithms due to low noise and microarchitectural leakages. The attack proposed by us depends on the algorithm and, therefore, does not depend much on such implementation-specific physical assumptions other than that of the fault. In other words, our attack remains unchanged for hardware implementations, provided one can realize the desired fault model, even if the masking is implemented very carefully with sufficient noise.}}  \\

%% file: sections/preliminaries.tex
\vspace{-15pt}
\section{Preliminaries}\label{sec:prelims}

\textbf{Notations: }The $k$-th bit of an integer $u \in \mathbb{Z}$ is denoted as \textcolor{black}{$u^{(k)}$}. $\mathbb{Z}_q$ is a ring of integer modulo $q$ and ${R}_q$ is a polynomial ring $\mathbb{Z}_q[X]/(X^{n}+1)$. ${R}_q^l$ is a ring containing vectors with $l$ polynomials of ${R}_q$ and we use ${R}_q^{l\times l}$ to represent a ring with $l\times l $ matrices over ${R}_q$. We denote a single polynomial with lower case letter (e.g., $x \in {R}_q$), a vector of polynomials with bold lower case letter (e.g., $\pmb{y} \in {R}_q^l$), and a matrix of polynomials with bold upper case letter (e.g., $\pmb{A} \in {R}_q^{l\times l}$). We denote the $i$-th ($i \in \{ 0,\ 1,\ \ldots,\ n-1 \}$) coefficient of a polynomial $x$ as $x[i]$ and the $i$-th ($i \in \{ 0,\ 1,\ \ldots,\ n-1 \}$) coefficient of $j$-th ($j \in \{ 0,\ 1,\ \ldots,\ l-1 \}$) polynomial in a vector of polynomials $\mathbf{y}$ as $\mathbf{y}[j][i]$. If an element $v$ is sampled from a set $S$ according to a distribution $\chi$, we present it by $v \leftarrow \chi(S)$. However, if the element $v$ is generated from a $seed_v$ with the help of a pseudorandom number generator and follows the distribution $\chi$ over the set $S$, then we denote it by $v \leftarrow \chi(S; seed_v)$. We use $\mathcal{U}$ as uniform distribution, and $\mathcal{\beta_\mu}$ as Centered Binomial Distribution (CBD) with standard deviation $\sqrt{\mu/2}$. $\xor$ denotes bit-wise XOR operation and $\&$ denotes bit-wise AND operation. $\lfloor w \rceil$ represents the rounding operation that outputs the closest integer of $w$ and is rounded upwards in case of ties (i.e., if $w = 1/2$ then $\lfloor w \rceil = 1$). $w \ll b$ and $w \gg b$ denotes logical shifting of $w$ by $b$ positions to left or right respectively. Here, $w$ is an integer, and $b$ is a natural number. These operations can be extended for a polynomial $x$ by performing them coefficient-wise (e.g., $x \gg b$ = $x[i] \gg b$, for each $i \in \{ 0,\ 1,\ \ldots,\ n-1 \}$). If $u\in\{0,1\}$ then $\overline{u} = 1 \xor u$. 

\subsection{LPR Public-Key Encryption}
The Ring-LWE (RLWE)~\cite{DBLP:conf/eurocrypt/LyubashevskyPR10} problem is a variant of the Regev's LWE~\cite{DBLP:journals/jacm/Regev09} problem introduced by Lyubashevsky et al. The decision version of RLWE problem states that if $a \leftarrow  \mathcal{U}(R_q)$, $s, e \leftarrow  \mathcal{\chi}(R_q)$, $b' \leftarrow  \mathcal{U}(R_q)$, and ${b} = {a} {s} + {e}$, then distinguishing (${a}$, ${b}$) and (${a}$, ${b}'$) is hard. The authors also proposed a public-key encryption scheme based on RLWE problem, which is commonly known as the LPR scheme. This scheme consists of three algorithms: (i) key generation ($\mathtt{LPR}{.}\mathtt{PKE}{.}\mathtt{KeyGen}$), (ii) encryption ($\mathtt{LPR}{.}\mathtt{PKE}{.}\mathtt{Enc}$), and (iii) decryption ($\mathtt{LPR}{.}\mathtt{PKE}{.}\mathtt{Dec}$), which are presented in~\autoref{fig:lprpke}. 
The key generation algorithm generates the public key and secret key pairs. In this algorithm, secret polynomial $s$ and the noise polynomial $e$ are sampled using a narrow distribution $\chi$ over the set $R_q$. The public polynomial $a$'s coefficients are drawn using a uniform distribution over the set $\mathbb{Z}_q$, and the other part of the public key is the polynomial $b$ is computed by computing $as + e$. The Encryption algorithm takes input as the public key $pk$ and message $m$ and produces the ciphertext pair $(u,\ v)$. Here, $u$ is the key containing part of the ciphertext and generated like $b$ in key generation. $v$ is the message containing components of the ciphertext. To generate $v$, an $n$-bit message $m$ is encoded to a message polynomial with mapping $\mathtt{Encode}$. One such mapping can be $\mathtt{Encode}: R_2 \rightarrow R_q$ defined by \textcolor{black}{$\mathtt{Encode}(m) = \sum_{i = 1}^n \lfloor q/2 \rceil m[i]$}, and then added with $bs' + e_2$. The decryption algorithm computes $m' = v-us$ to obtain a noisy version of the message $m$.
\textcolor{black}{ 
\begin{align*}
    m' & = v-us = bs'+e_2+\mathtt{Encode}(m)-(as'+e_1)s\\
    & = (as+e)s'+e_2+\mathtt{Encode}(m)-(as'+e_1)s = \mathtt{Encode}(m)+es'-e_1s+e_2 
\end{align*}
}
Here, $es'-e_1s+e_2$ is known as decryption noise. Also, $e,\ s',\ e_1,\ s,$ and $e_2$ are all sampled from a small distribution $\chi$. So, the decryption noise can be removed with a very high probability. This is performed by the $\mathtt{Decode}$ operation, which takes input as $m' \in R_q$ and outputs \textcolor{black}{$m \in {R}_2$ (shown in Figure \ref{fig:kyberfunc})}. 
\input{images/LPRPKE}

The chosen plaintext attack (CPA)-secure $\mathtt{LPR}{.}\mathtt{PKE}$ can be converted to a CCA (chosen ciphertext attack)-secure key encapsulation mechanism (KEM) using a post-quantum variant of the Fujisaki-Okamoto (FO) transformation proposed by Jiang et al.~\cite{Jiang2017}.

We call this FO transformation integrated CCA secure KEM as $\mathtt{LPR}{.}\mathtt{KEM}$, and present in Figure~\ref{fig:fokem}. $\mathcal{G},\ \mathcal{H}$, and $\mathtt{KDF}$ are three hash functions employed in this KEM as a part of FO transformation. This FO transformation is used in Kyber and Saber. 
\input{images/FOKEM}
\subsection{PQ KEM Schemes Kyber and Saber}\label{subsec:kyber}
Kyber is a $\mathtt{LPR}{.}\mathtt{KEM}$ like CCA secure KEM, but its security is based on the Module Learning with Errors (MLWE) problem. Here, the public polynomial of $\mathtt{LPR}{.}\mathtt{PKE}$ $a$ is a matrix of polynomials $\pmb{A} \in R^{l \times l}_q$, the secret polynomial $s$ is a vector of polynomials $\pmb{s} \in R^{l}_q$, and the error polynomial $e$ is a vector of polynomials $\pmb{e} \in R^{l}_q$. The MLWE problem states that if $\pmb{A} \leftarrow  \mathcal{U}(R^{l \times l}_q)$, $\pmb{s} \leftarrow  \mathcal{\beta_\mu}(R^{l}_q)$, $\pmb{e} \leftarrow  \mathcal{\beta_\mu}(R^{l}_q)$, $\pmb{b}' \leftarrow  \mathcal{U}(R^{l}_q)$, and $\pmb{b} = (\pmb{A} \pmb{s}) + \pmb{e}$, then ($\pmb{A}$, $\pmb{b}$) and ($\pmb{A}$, $\pmb{b}'$) are hard to distinguish. Kyber uses number-theoretic transformations to perform polynomial multiplication. Three security versions of Kyber named Kyber512, Kyber768, and Kyber1024, depending on different values of the parameter set, and are presented in Table~\ref{tab:Parameters_kyber}. The parameter set of Kyber contains prime modulus $q$ and two power-of-two moduli $p$ and $t$, which form the rings $R_q,\ R_p,\ R_t$. In the $\mathtt{Kyber}{.}\mathtt{PKE}{.}\mathtt{Enc}$, the $\mathtt{Compress}$ function shortens each coefficient of the ciphertext $c=(u,\ v)$. Each coefficient of $u$ is reduced from $R_q$ to $R_p$, and each coefficient of $v$ is reduced from $R_q$ to $R_t$. Conversely, the $\mathtt{Decompress}$ function in $\mathtt{Kyber}{.}\mathtt{PKE}{.}\mathtt{Dec}$, extend the ciphertext bits and maps each element of $R_p$ or $R_t$ to $R_q$. The $\mathtt{Encode}: \{0, 1\}^{n} \rightarrow R_q$ function converts $n$-bit message to a polynomial in $R_q$, and the $\mathtt{Decode}: R_q \rightarrow \{0, 1\}^{n} $ function reverse the effect of the $\mathtt{Encode}$ function.This is shown in \autoref{fig:kyberfunc}. Kyber uses two CBD distributions $\beta_{\eta_1}$ and $\beta_{\eta_2}$ to sample secret and error vector. Here, we brief the Kyber scheme and recommend \cite{Kyber-Kem} for additional details.
\input{images/functionKyberPKE}


\begin{table}[!ht]
\centering
\scriptsize
\caption{Parameters of Kyber with security and failure probability ~\cite{Kyber-Kem}}
\label{tab:Parameters_kyber}
\resizebox{0.90\textwidth}{!}{%
\begin{tabular}{|ll|ccccccc|c|c|c|}
\hline
\multicolumn{1}{|c}{\multirow{3}{*}{\begin{tabular}[c]{@{}c@{}}Scheme\\ Name\end{tabular}}} & \multirow{3}{*}{} & \multicolumn{7}{c|}{Parameters}                                                                                                                                                                                                                                                                   & \multirow{3}{*}{\begin{tabular}[c]{@{}c@{}}Post-quantum\\ Security\end{tabular}} & \multirow{3}{*}{\begin{tabular}[c]{@{}c@{}}Failure \\ Probability\end{tabular}} & \multirow{3}{*}{\begin{tabular}[c]{@{}c@{}}NIST\\ Security \\ Level\end{tabular}} \\ \cline{3-9}
\multicolumn{1}{|c}{}                                                                       &                   & \multicolumn{1}{c|}{\multirow{2}{*}{$l$}} & \multicolumn{1}{c|}{\multirow{2}{*}{$n$}}   & \multicolumn{1}{c|}{\multirow{2}{*}{$q$}}    & \multicolumn{1}{c|}{\multirow{2}{*}{$p$}} & \multicolumn{1}{c|}{\multirow{2}{*}{$t$}} & \multicolumn{1}{c|}{\multirow{2}{*}{$\eta_1$}} & \multirow{2}{*}{$\eta_2$} &                                                                                  &                                                                                 &                                                                                   \\
\multicolumn{1}{|c}{}                                                                       &                   & \multicolumn{1}{c|}{}                   & \multicolumn{1}{c|}{}                     & \multicolumn{1}{c|}{}                      & \multicolumn{1}{c|}{}                   & \multicolumn{1}{c|}{}                   & \multicolumn{1}{c|}{}                          &                           &                                                                                  &                                                                                 &                                                                                   \\ \hline
                                                                                            &                   & \multicolumn{1}{l|}{}                   & \multicolumn{1}{l|}{}                     & \multicolumn{1}{l|}{}                      & \multicolumn{1}{l|}{}                   & \multicolumn{1}{l|}{}                   & \multicolumn{1}{l|}{}                          & \multicolumn{1}{l|}{}     & \multicolumn{1}{l|}{}                                                            & \multicolumn{1}{l|}{}                                                           & \multicolumn{1}{l|}{}                                                             \\
Kyber512                                                                                    &                   & \multicolumn{1}{c|}{2}                  & \multicolumn{1}{c|}{\multirow{3}{*}{256}} & \multicolumn{1}{c|}{\multirow{3}{*}{3329}} & \multicolumn{1}{c|}{$2^{10}$}           & \multicolumn{1}{c|}{$2^{4}$}            & \multicolumn{1}{c|}{3}                         & 2                         & $2^{107}$                                                                        & $2^{-139}$                                                                      & 1                                                                                 \\
Kyber768                                                                                    &                   & \multicolumn{1}{c|}{3}                  & \multicolumn{1}{c|}{}                     & \multicolumn{1}{c|}{}                      & \multicolumn{1}{c|}{$2^{10}$}           & \multicolumn{1}{c|}{$2^{4}$}            & \multicolumn{1}{c|}{2}                         & 2                         & $2^{165}$                                                                        & $2^{-164}$                                                                      & 3                                                                                 \\
Kyber1024                                                                                   &                   & \multicolumn{1}{c|}{4}                  & \multicolumn{1}{c|}{}                     & \multicolumn{1}{c|}{}                      & \multicolumn{1}{c|}{$2^{11}$}           & \multicolumn{1}{c|}{$2^{5}$}            & \multicolumn{1}{c|}{2}                         & 2                         & $2^{232}$                                                                        & $2^{-174}$                                                                      & 5                                                                                 \\ \hline
\end{tabular}%
}
\end{table}
Saber is also a CCA secure KEM, and its security depends on the Module Learning with Rounding (MLWR) problem. The MLWR problem states that if $\pmb{A} \leftarrow  \mathcal{U}(R^{l \times l}_q)$, $\pmb{s} \leftarrow  \mathcal{\beta_\mu}(R^{l}_q)$, $\pmb{b}' \leftarrow  \mathcal{U}(R^{l}_p)$, $q > p$ and $\pmb{b} = \lfloor (q/p)(\pmb{A} \pmb{s}) \rceil$, then ($\pmb{A}$, $\pmb{b}$) and ($\pmb{A}$, $\pmb{b}'$) are hard to distinguish. Here the error vector of the MLWE problem is replaced by the rounding operation. The moduli $q,\ p,\ t$ that are used to form the rings $R_q,\ R_p,\ R_t$ in Saber algorithms are all power-of-two. The number of bits in each coefficient of a polynomial of the corresponding ring can be calculated as $\log_2(q) = \epsilon_q$, $\log_2(p) = \epsilon_p$, and $\log_2(t) = \epsilon_t$. 
The encoding and decoding operations of Saber are similar to $\mathtt{LPR}{.}\mathtt{PKE}$, and these operations are realized by using the shift operation due to the power-of-two moduli.
There are three security versions of Saber named LightSaber, Saber, and FireSaber, depending on different values of the parameter set, which is shown in Table~\ref{tab:Parameters_saber}. 
We suggest the original paper \cite{Saber_kem} for further specifics.
\begin{table}[!ht]
\scriptsize
\centering
\caption{Parameters of Saber with security and failure probability ~\cite{Saber_kem}}
\label{tab:Parameters_saber}
\resizebox{0.90\textwidth}{!}{%
\begin{tabular}{|ll|cccccc|c|c|c|}
\hline
\multicolumn{1}{|c}{\multirow{3}{*}{\begin{tabular}[c]{@{}c@{}}Scheme\\ Name\end{tabular}}} & \multirow{3}{*}{} & \multicolumn{6}{c|}{Parameters}                                                                                                                                                                                                                          & \multirow{3}{*}{\begin{tabular}[c]{@{}c@{}}Post-quantum\\ Security\end{tabular}} & \multirow{3}{*}{\begin{tabular}[c]{@{}c@{}}Failure \\ Probability\end{tabular}} & \multirow{3}{*}{\begin{tabular}[c]{@{}c@{}}NIST\\ Security \\ Level\end{tabular}} \\ \cline{3-8}
\multicolumn{1}{|c}{}                                                                       &                   & \multicolumn{1}{c|}{\multirow{2}{*}{$l$}} & \multicolumn{1}{c|}{\multirow{2}{*}{$n$}}   & \multicolumn{1}{c|}{\multirow{2}{*}{$q$}}        & \multicolumn{1}{c|}{\multirow{2}{*}{$p$}}        & \multicolumn{1}{c|}{\multirow{2}{*}{$t$}} & \multirow{2}{*}{$\mu$} &                                                                                  &                                                                                 &                                                                                   \\
\multicolumn{1}{|c}{}                                                                       &                   & \multicolumn{1}{c|}{}                   & \multicolumn{1}{c|}{}                     & \multicolumn{1}{c|}{}                          & \multicolumn{1}{c|}{}                          & \multicolumn{1}{c|}{}                   &                        &                                                                                  &                                                                                 &                                                                                   \\ \hline
                                                                                            &                   & \multicolumn{1}{l|}{}                   & \multicolumn{1}{l|}{}                     & \multicolumn{1}{l|}{}                          & \multicolumn{1}{l|}{}                          & \multicolumn{1}{l|}{}                   & \multicolumn{1}{l|}{}  & \multicolumn{1}{l|}{}                                                            & \multicolumn{1}{l|}{}                                                           & \multicolumn{1}{l|}{}                                                             \\
LightSaber                                                                                  &                   & \multicolumn{1}{c|}{2}                  & \multicolumn{1}{c|}{\multirow{3}{*}{256}} & \multicolumn{1}{c|}{\multirow{3}{*}{$2^{13}$}} & \multicolumn{1}{c|}{\multirow{3}{*}{$2^{10}$}} & \multicolumn{1}{c|}{$2^{3}$}            & 5                      & $2^{107}$                                                                        & $2^{-120}$                                                                      & 1                                                                                 \\
Saber                                                                                       &                   & \multicolumn{1}{c|}{3}                  & \multicolumn{1}{c|}{}                     & \multicolumn{1}{c|}{}                          & \multicolumn{1}{c|}{}                          & \multicolumn{1}{c|}{$2^{4}$}            & 4                      & $2^{172}$                                                                        & $2^{-136}$                                                                      & 3                                                                                 \\
FireSaber                                                                                   &                   & \multicolumn{1}{c|}{4}                  & \multicolumn{1}{c|}{}                     & \multicolumn{1}{c|}{}                          & \multicolumn{1}{c|}{}                          & \multicolumn{1}{c|}{$2^{6}$}            & 3                      & $2^{236}$                                                                        & $2^{-165}$                                                                      & 5                                                                                 \\ \hline
\end{tabular}%
}
\end{table}
\subsection{Masking Saber and Kyber}\label{subsec:masked-saber}

Due to many successful SCAs \cite{DBLP:conf/latincrypt/PesslP19,DBLP:journals/tches/KannwischerPP20,DBLP:journals/iacr/MujdeiBBKWV22,DBLP:journals/tches/RaviRCB20} on the decapsulation algorithm of the NIST KEM candidates, many masked implementations for those schemes have been proposed to prevent SCA \cite{FO-masked-saber,FO-masked-kyber-HeinzKLPSS22}.  In this work, we also target the masked decapsulation procedure of an LWE-based KEM. Therefore, we briefly describe the first-order masked decapsulation algorithm below, which can prevent first-order SCA.

The decapsulation of a general LWE-based KEM follows three steps: (i) decryption, (ii) re-encryption, and (iii) ciphertext comparison (\autoref{fig:fokem}). We must mask each of these three steps to protect the decapsulation algorithm from SCA. In first-order masking, the sensitive variable $w$ is split into two shares, $w_1$ and $w_2$. 
The relation between $w$ and $(w_1,\ w_2)$ for arithmetic masking is $(w_1 + w_2) \bmod{q}= w$ and for Boolean masking it is $w_1 \xor w_2 = w$.
LWE-based KEMs primarily utilize linear polynomial arithmetic operations over a ring $R_q$, such as ring multiplication with one masked and another unmasked input, ring addition, and ring subtraction. These operations are duplicated and performed on each share independently in masked settings. For masking them, arithmetic masking techniques are used. The decapsulation also has non-linear arithmetic over $R_q$. Those are shift operations on polynomials, and hash functions $\mathcal{H},\ \mathcal{G}$, the extension function used for sampling pseudorandom numbers $\mathtt{XOF}$, the CBD $\beta_\mu$, and ciphertext equality checking operation $=?$. These operations are expressed in bit operations and masked with Boolean masking. Except for these operations, the masked decapsulation needs two share conversion algorithms \textit{i.e.} arithmetic to Boolean (\texttt{A2B}) and Boolean to arithmetic (\texttt{B2A}) conversions. The \texttt{B2A} conversion occurs inside the masked CBD ($\beta_\mu$).

In Saber's decapsulation operation, the ring is $R_p$ instead of $R_q$. 
A pictorial version of this masked decapsulation operation of Saber is presented in \autoref{fig:sabermaskeddecaps}. 
In this work, we will focus on the part surrounded by the red rectangle in the figure, which includes the \texttt{A2B} conversion and shift operations. A version of \texttt{A2B} conversion used in Saber is shown in Algorithm \ref{algo:A2B}.

\input{images/sabermasked}

\begin{algorithm}[!ht]
\caption{ First-order masked arithmetic to Boolean conversion (\texttt{A2B})~\cite{Coron2014_A2B} }
\label{algo:A2B}

\Input{$(a_1,\ a_2)\in R_{2^k} \times R_{2^k}$ such that $(a_1 + a_2) \bmod{2^k}= {w} \in R_{2^k}$}
\Output{$(b_1,\ b_2)\in R_{2^k} \times R_{2^k}$ such that $b_1 \xor b_2 = w$}
\BlankLine
$u_1\leftarrow \mathcal{U}(R_{2^k})$;    $u_2 = u_1 \xor a_1$ \\
$v_1\leftarrow \mathcal{U}(R_{2^k})$;    $v_2 = v_1 \xor a_2$ \\
$(b_1,\ b_2) = \mathtt{SecAdd} ((u_1,\ u_2),\ (v_1,\ v_2),\ k)$ [Algorithm \ref{algo:SecAdd}]\\
\algorithmicreturn {$(b_1,\ b_2)$}
\end{algorithm}
\begin{algorithm}[!ht]
\caption{ SecAdd~\cite{Coron2014_A2B} }
\label{algo:SecAdd}

\Input{$(u_1,\ u_2),\ (v_1,\ v_2)\in R_{2^k} \times R_{2^k}$ such that $ u_1 \xor u_2 = \hat{u}$ and $ v_1 \xor v_2 = \hat{v}$}
\Output{$(z_1,\ z_2)\in R_{2^k} \times R_{2^k}$ such that $ z_1 \xor z_2 = (\hat{u}+\hat{v}) \bmod{2^k}$}
\BlankLine
$(c_1^1,\ c_2^1) = (0,\ 0)$\\
$r_0,\ r_1,\ r_2 \leftarrow \mathcal{U}(R_{2^k})$\\
\For{\texttt{i=1 to k-1 }}
{
    $tmp1_1 = (u_1^i \& v_1^i) \xor r_0^i$;    $tmp1_2 = (u_2^i \& v_2^i) \xor r_0^i$\\
    $tmp2_1 = (u_1^i \& c_1^i) \xor r_1^i$;    $tmp2_2 = (u_2^i \& c_2^i) \xor r_1^i$\\
    $tmp3_1 = (v_1^i \& c_1^i) \xor r_2^i$;    $tmp3_2 = (v_2^i \& c_2^i) \xor r_2^i$\\
    $c_1^{i+1} = tmp1_1 \xor tmp2_1 \xor tmp3_1$;    $c_2^{i+1} = tmp1_2 \xor tmp2_2 \xor tmp3_2$\\
}
$z_1 = u_1 \xor v_1 \xor c_1$;    $z_2 = u_2 \xor v_2 \xor c_2$\\
\algorithmicreturn{ $(z_1,\ z_2)$}
\end{algorithm}

\begin{figure}[!tbp]
  \centering
  {\includegraphics[width=0.7\textwidth]{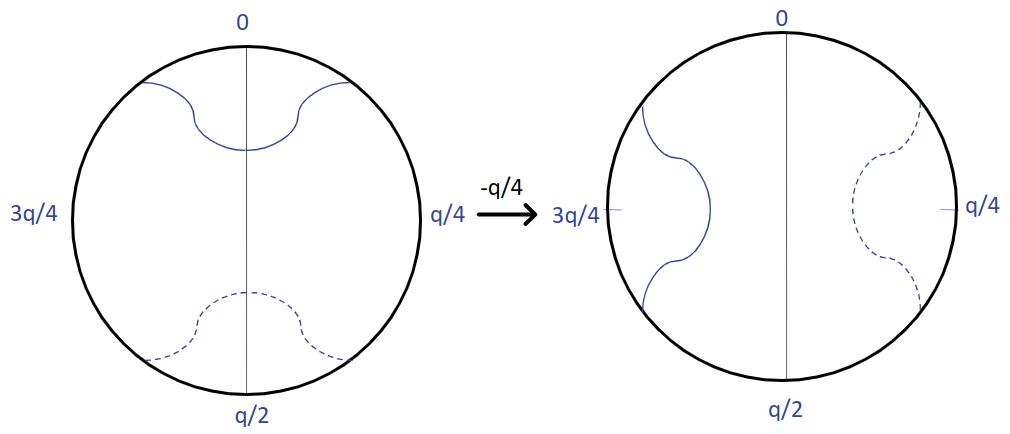}}
  \hfill
  {\includegraphics[width=0.85\textwidth]{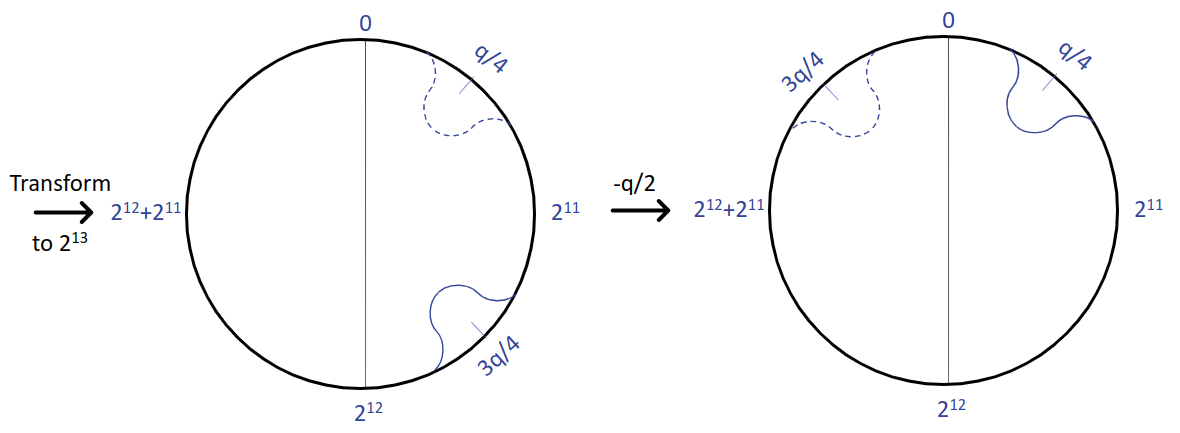}}
  \caption{\textcolor{black}{The steps of the masked decoding algorithm of Kyber presented in Algorithm~\ref{algo:masked-decode}\cite{1st-order_masked_comparator}.}}
  \label{fig:masked-decoding}
\end{figure}

In Kyber, the ring modulus $q$ is prime, the share conversion algorithms \texttt{A2B} (\texttt{A2B}$_q$) and \texttt{B2A} (\texttt{B2A}$_q$) are different compared to Saber, which uses power-of-two modulus. In Kyber, the shares are first transferred from prime to a power-of-two modulus, and the aforementioned \texttt{A2B} conversion (Algorithm \ref{algo:A2B}) or a \texttt{B2A} conversion algorithm with input in power-of-two modulus is performed. For this reason, a couple of extra masked modules, such as $\mathtt{Decode}$, $\mathtt{Encode}$, $\mathtt{Compress}$, and \texttt{transform-power-of-2} are required in masked Kyber. In Kyber, the \texttt{A2B} conversions are used inside the $\mathtt{Decode}$ and $\mathtt{Comp}$ components, and the \texttt{B2A} conversions are inside the $\mathtt{Encode}$. \autoref{fig:masked-decoding} is an illustrated version of the first-order masked decapsulation algorithm of Kyber. The function we exploit in this work is the \texttt{A2B} conversion in the $\mathtt{Decode}$ module, which is surrounded by the red rectangle in the figure. 
\textcolor{black}{We present a first-order masked $\mathtt{Decode}$ procedure in Algorithm \ref{algo:masked-decode}.}
\textcolor{black}{According to this algorithm, $(m'_{a_1},\, m'_{a_2})$ are arithmetically masked shares of the input of the $\mathtt{Decode}$ operation $m'$. We use Figure \ref{fig:masked-decoding} to illustrate the steps of the masked decoding algorithm for a coefficient of $m'$, say $m'[i]$. Here, the first figure shows the distribution of a coefficient of $m'[i]$. The second figure presents the distribution of that coefficient after the subtraction with $q/4$. The third figure expresses the distribution of that coefficient after the transformation from $\mathbb{Z}_{q}$ to $\mathbb{Z}_{2^{k+1}}$, where $k = \lceil\log_2(q)\rceil$. The last figure displays the distribution of that coefficient after the subtraction with $q/2$. After this, the \texttt{A2B} conversion takes place and produces Boolean shares $(d_{b_1}[i],\ d_{b_2}[i])$ in Algorithm \ref{algo:masked-decode}. Then the most significant bits (MSBs) of $(d_{b_1}[i],\ d_{b_2}[i])$ are calculated by applying $\mathtt{MSB}$ coefficient-wise. $(d_{b_1}[i],\ d_{b_2}[i])$ act as the Boolean shares of $m[i] = (m_{b_1}[i],\ m_{b_2}[i])$. Here, $k$th and $(k+1)$th bit of $(d_{b_1}[i],\ d_{b_2}[i])$ both can serve as the Boolean shares of the message bit $m[i]$. Therefore,
\begin{equation}\label{eq:k=k+1}
    (d_{b_1}[i]^{(k+1)},\ d_{b_2}[i]^{(k+1)}) = (d_{b_1}[i]^{(k)},\ d_{b_2}[i]^{(k)}) = (m_{b_1}[i],\ m_{b_2}[i])
\end{equation}
All other bits of $(d_{b_1}[i],\ d_{b_2}[i])$ act as decryption noise. As noted in Section \ref{sec:intro}, the decryption noise is linearly related to the secret key.  
}
\vspace{-10pt}
\begin{algorithm}[!ht]
\caption{ First-order masked decode algorithm~\cite{1st-order_masked_comparator} }
\label{algo:masked-decode}
\Input{$(m'_{a_1},\ m'_{a_2})\in R_q \times R_q$ such that $m' = (m'_{a_1} + m'_{a_2}) \bmod{q}$, and $k = \lceil\log_2(q)\rceil$}
\Output{$(m_{b_1},\ m_{b_2})\in R_2\times R_2$ where $m = m_{b_1} \xor m_{b_2} = \mathtt{Decode}((m'_{a_1} + m'_{a_2}) \bmod{q})$}
\BlankLine
$m'_{a_1} = m'_{a_1} - \lfloor q/4 \rfloor$ \\
\tcc{\text{transfers arithmetic shares from }$\bmod{q}$\text{ to }$\bmod{2^{k+1}}$}
$(g_{a_1},\ g_{a_2}) = \texttt{transform-power-of-2}(m'_{a_1},\ m'_{a_2},\ (k+1))$ \\
$g_{a_1} = g_{a_1} - \lfloor q/2 \rfloor$ \\
$(d_{b_1},\ d_{b_2}) = \texttt{A2B} ((g_{a_1},\ g_{a_2}),\ (k+1))$ [Algorithm \ref{algo:A2B}]\\
$m_{b_1} = \mathtt{MSB}(d_{b_1});\  m_{b_2} = \mathtt{MSB}(d_{b_2})$\\
\algorithmicreturn {$(m_{b_1},\ m_{b_2})$}
\end{algorithm}

%% file: images/LPRPKE.tex
\begin{figure}[!ht]
\centering
\fbox{\begin{varwidth}{0.90\textwidth}
\begin{subfigure}[t]{0.47\textwidth}
    \raggedright 
    \begin{small}
    $\mathtt{LPR}{.}\mathtt{PKE}{.}\mathtt{KeyGen} ()$
    \begin{enumerate}[wide=0em, itemsep=0pt, parsep=0pt, font=\scriptsize\tt\color{gray}]
            \item $a \leftarrow  \mathcal{U}(R_q) $ \\
            \item $s,\ e \leftarrow  \chi( R_q)$ \\
            \item $b =  (as + e) \in R^{l}_q$ \\
            \item \textbf{return} $(pk = (a,\ b),\ sk = (a,\ s))$
    \end{enumerate} 
    \end{small}
\end{subfigure}
\vspace{3mm}
\begin{subfigure}[t]{0.52\textwidth}
    \raggedright
    \begin{small}
    $\mathtt{LPR}{.}\mathtt{PKE}{.}\mathtt{Enc} (pk = (a,\ {b}),\ \texttt{message}\ m \in \mathbb{Z}_2^n)$
    \begin{enumerate}[wide=0em, itemsep=0pt, parsep=0pt, font=\scriptsize\tt\color{gray}]
        \item $s',\ e_1,\ e_2 \leftarrow  \chi(R_q)	$\\
        \item $u = ( a{s'} + {e_1}) \in R_q$\\
        \item $v = b{s'} + {e_2} + \mathtt{Encode}(m)$\\ 
        \item \textbf{return} $c =(u,\ v)$
    \end{enumerate} 
    \end{small}
\end{subfigure}
\begin{subfigure}[t]{\textwidth}
    \raggedright
    \begin{small}
    $\mathtt{LPR}{.}\mathtt{PKE}{.}\mathtt{Dec}(sk= (a,\ s),\ c =(u,\ v))$
    \begin{enumerate}[wide=0em, itemsep=0pt, parsep=0pt, font=\scriptsize\tt\color{gray}]
        \item $m' = v - us \in R_q$ \\
        \item $m = \mathtt{Decode}(m') \in R_2$  \\
        \item \textbf{return} $m$
    \end{enumerate} 
    \end{small}
\end{subfigure}
\end{varwidth}}
\caption{$\mathtt{LPR}{.}\mathtt{PKE}$~\cite{DBLP:conf/eurocrypt/LyubashevskyPR10}}
\label{fig:lprpke}
\end{figure}

%% file: images/FOKEM.tex
\begin{figure}[!ht]
\centering
\fbox{\begin{varwidth}{0.90\textwidth}
\begin{subfigure}[t]{0.47\textwidth}
    \begin{small}
    \raggedright 
    $\mathtt{LPR}{.}\mathtt{KEM}{.}\mathtt{KeyGen} ()$
    \begin{enumerate}[wide=0em, itemsep=0pt, parsep=0pt, font=\scriptsize\tt\color{gray}]
        \item $(pk,\ sk)  = \mathtt{LPR}{.}\mathtt{PKE}{.}\mathtt{KeyGen} ()$ \\
        \item $pkh = \mathcal{H}(pk) $ \\
        \item $z \leftarrow \mathcal{U}(\{0,\ 1\}^{n}$) \\
        \item \textbf{return} $(pk,\ \widetilde{sk} =  (sk,\ z,\ pkh))$
    \end{enumerate} 
    \end{small}
\end{subfigure}
\begin{subfigure}[t]{0.47\textwidth}
    \begin{small}
    \raggedright
    $\mathtt{LPR}{.}\mathtt{KEM}{.}\mathtt{Encaps} (pk)$
    \begin{enumerate}[wide=0em, itemsep=0pt, parsep=0pt, font=\scriptsize\tt\color{gray}]
        \item $m  \leftarrow  \mathcal{U}(\{0,\ 1\}^{n}) $ \\
        \item $m  = \mathcal{H}(m)$ \\
        \item $(\hat{K},\ r) = \mathcal{G}(\mathcal{H}(pk), m)$ \\
        \item $c = \mathtt{LPR}{.}\mathtt{PKE}{.}\mathtt{Enc} (pk,\ m;\ r)$ \\
        \item $K = \mathtt{KDF}(\hat{K},\ \mathcal{H}(c))$
        \item \textbf{return} $(c,\ K)$
    \end{enumerate} 
    \end{small}
\end{subfigure}
\begin{subfigure}[t]{\textwidth}
    \vspace{-2mm}
    \begin{small}
    \raggedright
    $\mathtt{LPR}{.}\mathtt{KEM}{.}\mathtt{Decaps} (\widetilde{sk} = (sk,\ z,\ pkh),\ pk,\ c)$
    \begin{enumerate}[wide=0em, itemsep=0pt, parsep=0pt, font=\scriptsize\tt\color{gray}]
        \item $m  =  \mathtt{LPR}{.}\mathtt{PKE}{.}\mathtt{Dec} (sk,\ c )$ \\
        \item $(\hat{K}',\ r') = \mathcal{G}(pkh,\ m)$ \\
        \item $c_* = \mathtt{LPR}{.}\mathtt{PKE}{.}\mathtt{Enc} (pk,\ m;\ r')$ \\
        \item \textbf{if: } $c=c_*$
        \item \quad \textbf{return} $ K = \mathtt{KDF}(\hat{K}',\ \mathcal{H}(c))$
        \item \textbf{else: }
        \item \quad \textbf{return} $ K = \mathtt{KDF}(z,\ \mathcal{H}(c))$  
    \end{enumerate} 
    \end{small}
\end{subfigure}
\end{varwidth}}
\caption{CCA secure $\mathtt{LPR}{.}\KEM$~\cite{Jiang2017}}
\label{fig:fokem}
\end{figure}

%% file: images/functionKyberPKE.tex
\begin{figure}[ht]
\centering
\fbox{\begin{varwidth}{0.90\textwidth}
\begin{subfigure}[t]{0.40\textwidth}
    \begin{small}
    \raggedright 
    $\mathtt{Compress} (v',\ t) $ (referred as $\mathtt{Comp}$)
    \begin{enumerate}[wide=0em, itemsep=0pt, parsep=0pt, font=\scriptsize\tt\color{gray}]
        \item $v  = \lfloor (t/q) v' \rceil \bmod{t}$ \\
        \item \textbf{return} $(v)$
    \end{enumerate} 
    \end{small}
    \vspace{2mm}
\end{subfigure}
\begin{subfigure}[t]{0.58\textwidth}
    \begin{small}
    \raggedright 
    $\mathtt{Decompress} (v,\ t)$ (referred as $\mathtt{Decomp}$)
    \begin{enumerate}[wide=0em, itemsep=0pt, parsep=0pt, font=\scriptsize\tt\color{gray}]
        \item $v'  = \lfloor (q/t) v \rceil $ \\
        \item \textbf{return} $(v')$
    \end{enumerate} 
    \end{small}
    \vspace{2mm}
\end{subfigure}
\begin{subfigure}[t]{0.40\textwidth}
    \begin{small}
    \raggedright 
    $\mathtt{Encode} ({m})$
    \begin{enumerate}[wide=0em, itemsep=0pt, parsep=0pt, font=\scriptsize\tt\color{gray}]
        \item {for} {\texttt{i=1 to n}}:
        \item  \text{     } $m'[i] = \lfloor q/2 \rceil {m}[{i}] \in \mathbb{Z}_q$ \\
        \item \textbf{return} $(m'  \in R_q)$
    \end{enumerate} 
    \end{small}
\end{subfigure}
\begin{subfigure}[t]{0.58\textwidth}
    \begin{small}
    \raggedright 
    $\mathtt{Decode} (m')$
    \begin{enumerate}[wide=0em, itemsep=0pt, parsep=0pt, font=\scriptsize\tt\color{gray}]
        \item {for} {\texttt{i=1 to n}}:
        \item  \text{     } ${m}[{i}] = (((m'[i] \ll 1) + \lfloor q/2 \rceil) / q ) \& 1 \in \mathbb{Z}_2$ \\
        \item \textbf{return} $({m} \in \{0,1\}^{n})$
    \end{enumerate} 
    \end{small}
\end{subfigure}
\end{varwidth}}
\caption{$\mathtt{Compress}$, $\mathtt{Decompress}$, $\mathtt{Encode}$, and $\mathtt{Decode}$ functions of $\mathtt{Kyber}$~\cite{Kyber-Kem}}
\label{fig:kyberfunc}
\end{figure}

%% file: images/sabermasked.tex
\begin{figure}[!ht]
    \centering
\setlength{\nodedistance}{4mm}
\resizebox{0.75\textwidth}{!}{
\begin{tikzpicture}[ 
node distance = \nodedistance,
operation/.style = {circle, draw=black},
bigoperation/.style = {rounded corners, draw=black},
operationsecret/.style = {circle, draw=black, fill=black!20!white, text=black, very thick},
bigoperationsecret/.style = {rounded corners, draw=black, fill=black!20!white, text=black, very thick},
Bigoperationsecret/.style = {rounded corners, draw=black, fill=black!20!white, text=black, very thick, minimum height=4\nodedistance, , minimum width=2\nodedistance},
operant/.style = {},
halfblockdraw/.style = {draw, rounded corners},
line/.style = {draw, -Latex},
connect with angle/.style=
{to path={let \p1=(\tikztostart), 
\p2=(\tikztotarget) in -- ++(0, {(\x2-\x1)*tan(#1-90)-(\y1-\y2)}) -- (\tikztotarget)}},
connect with angle hor/.style=
{to path={let \p1=(\tikztostart), 
\p2=(\tikztotarget) in -- ++({(\x2-\x1)-(\y1-\y2)*tan(#1-90)}, 0) -- (\tikztotarget)}},
                       ]
\node (multbps) [operationsecret]  {X};
\node (bp) [operant, left=of multbps.west]  {$\pmb{u}$};
\draw[->] (bp) -- (multbps);

\node (s) [operant, above=of multbps.north]  {$\pmb{s}_1$};
\draw[->, very thick] (s) -- (multbps);

\node (multbps2) [operationsecret, below=of multbps]  {X};
\draw[->] (bp) -- (bp |- multbps2) -- (multbps2);

\node (s) [operant, below=of multbps2]  {$\pmb{s}_2$};
\draw[->, very thick] (s) -- (multbps2);

\node (addvh) [operationsecret, right=of multbps]  {$+$};
\draw[->, very thick] (multbps.east) -- (addvh.west);
\node (h2) [operant, above=of addvh.north]  {$\pmb{h}_2$};
\draw[->] (h2) -- (addvh);

\node (cm) [operant, below=6\nodedistance of bp]  {$v$};
\node (shiftcm) [bigoperation, right=of cm]  {$\ll$};
\draw[->] (cm) -- (shiftcm);

\coordinate [right=of addvh](addvhright);
\node (subcmv) [operationsecret, right=of multbps2] {$-$};
\draw[->, very thick] (multbps2) -- (subcmv);
\draw[->] (shiftcm) -- (subcmv |- shiftcm) -- (subcmv);

\node (tobin) [Bigoperationsecret, right=of $(subcmv.east)!0.5!(addvh.east)$]  {$A2B$};
\draw[->, very thick] (addvh) -- (tobin.west |- addvh);
\draw[->, very thick] (subcmv) -- (tobin.west |- subcmv);


\node (shift1) [bigoperationsecret, right=of $(tobin.east |- addvh)$]  {$\gg$};
\node (shift2) [bigoperationsecret, right=of $(tobin.east |- subcmv)$]  {$\gg$};
\draw[->, very thick] (tobin.east |- addvh) -- (shift1);
\draw[->, very thick] (tobin.east |- subcmv) -- (shift2);

\node[draw=red, -, very thick, fit=(tobin) (shift1) (shift2)] {};

\node (g) [Bigoperationsecret, right= 2\nodedistance of $(shift1.east)!0.5!(shift2.east)$]  {$\mathcal{G}$};
\node (pkh) [operant, above=of g]  {$pkh$};
\coordinate  (m1) at ($(shift1.east)!0.6!(g.west |- addvh)$);
\coordinate  (m2) at ($(shift2.east)!0.3!(g.west |- subcmv)$);
\draw[->, transform canvas={xshift=0.5em}] (pkh) -- (g);
\draw[->, transform canvas={xshift=-0.5em}] (pkh) -- (g);
\draw[->, very thick] (shift1.east) -- (g.west |- addvh);
\draw[->, very thick] (shift2.east) -- (g.west |- subcmv);


\node (K) [operant, below=of g]  {$\hat{K}'$};
\draw[->, very thick, transform canvas={xshift=0.5em}] (g) -- (K);
\draw[->, very thick, transform canvas={xshift=-0.5em}] (g) -- (K);

\node (xof) [Bigoperationsecret, right=of g]  {$XOF$};
\draw[->, very thick] (g.east |- addvh) -- (xof.west |- addvh);
\draw[->, very thick] (g.east |- subcmv) -- (xof.west |- subcmv);

\node (beta) [Bigoperationsecret, right=of xof]  {$\beta_{\mu}$};
\draw[->, very thick] (xof.east |- addvh) -- (beta.west |- addvh);
\draw[->, very thick] (xof.east |- subcmv) -- (beta.west |- subcmv);

\coordinate (betaout1) at (beta.east |- addvh);
\coordinate (betaout2) at (beta.east |- subcmv);

\node (multasp) [operationsecret, right= 3\nodedistance of betaout1]  {X};
\node (multasp2) [operationsecret, above=of multasp]  {X};

\node (gena) [bigoperation, above left=2\nodedistance and \nodedistance of multasp2]  {$\mathcal{U}$};
\draw[->] (gena) edge[connect with angle=135] (multasp2);
\draw[->] (gena) edge[connect with angle=135] (multasp);


\node (seeda) [operant, left=of gena]  {$\seeda$};
\draw[->] (seeda) -- (gena);

\node (multbsp) [operationsecret, below=of multasp]  {X};
\node (multbsp2) [operationsecret, below=of multbsp]  {X};

\node (b) [operant,  below left=2\nodedistance and \nodedistance of multbsp2]  {$\pmb{b}$};
\draw[->] (b) edge[connect with angle=45] (multbsp);
\draw[->] (b) edge[connect with angle=45] (multbsp2);

\coordinate  (r1) at ($(betaout1)!0.2!(multasp)$);
\coordinate  (r2) at ($(betaout2)!0.6!(multasp)$);

\draw[->, very thick]  (betaout1) -- (multbsp); 
\draw[->, very thick]  (betaout1) -- (multasp2); 

\draw[->, very thick]  (betaout2) -- (multbsp2); 
\draw[->, very thick]  (betaout2) -- (multasp); 

\node (addh1) [operationsecret, right=of multasp2]  {$+$};
\draw[->, very thick] (multasp2) -- (addh1);
\node (h1) [operant, above=of addh1]  {$\pmb{h}$};
\draw[->] (h1) -- (addh1);

\node (addh2) [operationsecret, right=of multbsp2]  {$+$};
\draw[->, very thick] (multbsp2) -- (addh2);
\node (h) [operant, below=of addh2]  {$h_1$};
\draw[->] (h) -- (addh2);

\node (addm) [operationsecret, right=of addh2]  {$+$};
\node (addm1) [operationsecret] at (addm |- multbsp)  {$+$};
\coordinate [below=0.9\nodedistance of b](tmp2);
\coordinate [below=0.2\nodedistance of b](tmp1); 
\draw[->, very thick] (m2) -- (m2 |- tmp2) -- (addm |- tmp2) -- (addm);
\coordinate (iammakingthiscodeugly) at ($(addm)!0.5!(addh2)$);
\draw[very thick] (m1) -- (m1 |- tmp1) -- (iammakingthiscodeugly |- tmp1) edge[->,connect with angle=45] (addm1);

\draw[->, very thick] (multbsp) -- (addm1);
\draw[->, very thick] (addh2) -- (addm);

\coordinate[right= 2\nodedistance of addm] (dividex);
\coordinate (dividey1) at ($(multasp)!0.5!(multasp2)$);
\coordinate (dividey2) at ($(multbsp)!0.5!(multbsp2)$);
\node (divide2) [Bigoperationsecret] at (dividex |- dividey1)  {$A2B$};
\node (divide1) [Bigoperationsecret] at (dividex |- dividey2)  {$A2B$};
\draw[->, very thick] (addm) -- (divide1.west |- addm);
\draw[->, very thick] (addh1) -- (divide2.west |- addh1);
\draw[->, very thick] (addm1) -- (divide1.west |- addm1);
\draw[->, very thick] (multasp) -- (divide2.west |- multasp);

\node (shift3) [bigoperationsecret, right=of $(divide2.east |- addh1)$]  {$\gg$};
\node (shift4) [bigoperationsecret, right=of $(divide2.east |- multasp)$]  {$\gg$};
\node (shift5) [bigoperationsecret, right=of $(divide1.east |- addm1)$]  {$\gg$};
\node (shift6) [bigoperationsecret, right=of $(divide1.east |- addm)$]  {$\gg$};
\draw[->, very thick] (divide2.east |- addh1) -- (shift3);
\draw[->, very thick] (divide2.east |- multasp) -- (shift4);
\draw[->, very thick] (divide1.east |- addm1) -- (shift5);
\draw[->, very thick] (divide1.east |- addm) -- (shift6);

\coordinate[right= 2\nodedistance of divide1] (output);
\node (bp2) [operant, right=of $(shift3.east)$] {$\pmb{u}_{*1}$};
\node (bp22) [operant, right=of $(shift4.east)$]  {$\pmb{u}_{*2}$};
\node (cm22) [operant, right=of $(shift5.east)$]  {${v}_{*1}$};
\node (cm2) [operant, right=of $(shift6.east)$]  {${v}_{*2}$};
\draw[->, very thick] (shift3.east |- bp2) -- (bp2);
\draw[->, very thick] (shift4.east |- bp22) -- (bp22);
\draw[->, very thick] (shift5.east |- cm22) -- (cm22);
\draw[->, very thick] (shift6.east |- cm2) -- (cm2);


\node[draw=black, dotted, thick, fit=(bp2) (cm2)](ciphertext2) {};
\node[draw=black, dotted, thick, fit=(bp) (cm)](ciphertext) {};
\coordinate [below=1.5\nodedistance of tmp2](x); 
\coordinate (y) at ($(ciphertext)!0.5!(ciphertext2)$);
\node (comp) [operationsecret, very thick] at (y |- x)  {$=?$};

\draw[->] (ciphertext) -- (ciphertext |- x) -- (comp);
\draw[->, very thick] (ciphertext2) -- (ciphertext2 |- x) -- (comp);

\node (correct) [bigoperation, below left=2\nodedistance and 2\nodedistance of comp]  {return $\mathcal{H}(\hat{K}', c)$};
\draw[->] (comp.south) -- (correct);
\node[operant,  below left=0.5\nodedistance and 1.3\nodedistance of comp] (yes) {yes};

\node (correct) [bigoperation, below right=2\nodedistance and 2\nodedistance of comp]  {return $\mathcal{H}(z, c)$};
\draw[->] (comp.south) -- (correct);
\node[operant,  below right=0.5\nodedistance and 1.3\nodedistance of comp] (no) {no};
\end{tikzpicture}}

\caption{First-order masked $\mathtt{Saber}.\mathtt{KEM}.\mathtt{Decaps}$ algorithm \cite{FO-masked-saber}. The highlighted operations in color gray are influenced by the non-ephemeral secret-key $\pmb{s}$ and use masking to prevent SCA. The component we focus on in this work is enveloped with the red rectangle.}

\label{fig:sabermaskeddecaps}
\end{figure}

%% file: sections/fault_attack.tex
\vspace{-5pt}
\section{Fault Propagation through the Carry Chain}\label{sec:fault-attack}
\textcolor{black}{In this section, we present our main observation, which results in key recovery fault attacks on the masked decapsulation algorithms of certain CCA secure LWE-based KEMs. 
Broadly, our attack exploits the $\mathtt{Decode}$ component (ref. Figure~\ref{fig:maskeddecaps}) that removes the decryption noise and produces the message in the decapsulation algorithm. However, unlike~\cite{PesslP21}, which skips the first subtraction operation (line 1. in Algorithm~\ref{algo:masked-decode}) in the $\mathtt{Decode}$ module, we target the $\mathtt{A2B}$ algorithm. The \texttt{A2B} adds the arithmetic shares using Boolean masking and generates Boolean shares amenable to the shift operation so that the message bits can be extracted easily. More precisely, we target one of the arithmetic shares input to this module with faults. For simplicity, we limit our discussion to two shares (ref. Algorithm~\ref{algo:masked-decode}) in the next few paragraphs, and later explain the many shares case. Moreover, for ease of explanation, we limit the discussion to the $\mathtt{Decode}$ module of Kyber only (described in the last section). The differences for the Saber case will be explained at the end.}

\textcolor{black}{The output of the $\mathtt{Decode}$ stage are the message bits (masked), which are inputs to the re-encryption stage. As described in the last section, these (masked) message bits are derived from the (masked) message coefficients containing (masked) noise. \emph{The faults induced by us at the input of the $\mathtt{A2B}$ (at a specific bit of any one of the shares) propagate to these decoded (masked) message bits, conditioned on a specific (unmasked) bit of the noise.} In other words, whether or not the fault propagates to the (masked) message bits (and, therefore, causes a decryption failure) depends on the actual unmasked value of some noise bit. Now the noise is dependent on the secret key $\pmb{s}$. Therefore, based on this fault propagation property, we can construct inequalities (each corresponding to one ciphertext) involving the secret, and solve a system of such inequalities using the Belief Propagation (BP) algorithm. The BP algorithm finally returns the long-term secret key. Note that, in this attack, we always work with valid ciphertexts, \textit{i.e.} the ciphertexts that will be generated using proper execution of the encapsulation mechanism of the scheme.
So, the probability of decryption failure (without fault) during the decapsulation procedure is identical to the preassigned decryption failure probability of the scheme, which is very small ($<2^{-128}$). Therefore, we can safely assume that all the observed decryption failures happen due to the induced faults. Next, we explain the information leakage that we exploit due to the fault propagation through carry chains of the $\mathtt{A2B}$.   
}

\input{images/masked_scheme}



\subsection{Fault Propagation through First-order Masked \texttt{A2B}} \label{subsec:fault-propagation}

\textcolor{black}{
We begin our discussion with stuck-at faults. Without loss of generality, suppose we introduce \texttt{stuck-at-1} bit-fault (on an input share) at the $(k-1)$th bit of the $i$-th input coefficient (\textit{i.e.} $g_{a_1}[i]^{(k-1)}$) to the \texttt{A2B} conversion. Here $i \in \{0,\ 1,\ \ldots,\ n-1\}$\footnote{\textcolor{black}{Please note that we start counting of the index $k$ from $1$.}}. Let $g = (g_{a_1} + g_{a_2}) \bmod{2^{k+1}} = (d_{b_1} \xor d_{b_2})$. Here, $g[i]^{(k+1)} = (d_{b_1}[i]^{(k+1)} \xor d_{b_2}[i]^{(k+1)}) = (m_{b_1}[i] \xor m_{b_2}[i]) = m[i]$ ($m[i]$s are the bits of the message).
Henceforth in this section, we will only focus on a single coefficient. So, for the sake of simplicity, we rename the following:
\begin{itemize}
    \item $g_{a_1}[i]$ as $x$, $g_{a_2}[i]$ as $y$, and $g[i]$ as $z$. \textcolor{black}{These are $(k+1)$ bit registers}.
    \item $m_{b_1}[i]$ as $m_1$, $m_{b_2}[i]$ as $m_2$, and $m[i]$ as $\hat{m}$. \textcolor{black}{These are $1$ bit registers}.
\end{itemize}
Let us assume after the application of the \texttt{stuck-at-1} fault at $x^{(k-1)}$, $x$ becomes $x^*$, $z = (x + y) \bmod{2^{(k+1)}}$ becomes ${z}^*$, and the message bit $\hat{m}$ becomes $\hat{m}^*$. 
We introduce the following two events: 
\begin{itemize}
    \item \textbf{Fault activation}: The injected \texttt{stuck-at-1} fault at $x^{(k-1)}$ is active if $x^{(k-1)} = 0$, else it is inactive. If the fault is active, then $x^{*(k-1)} = \overline{x^{(k-1)}}$ (the value of the bit changes from 0 to 1).
    \item \textbf{Fault propagation}: The injected \texttt{stuck-at-1} fault at $x^{(k-1)}$ propagates to $z^{(k)}$ if and only if $z^{*(k)} = \overline{z^{(k)}}$. The injected fault propagates to $z^{(k+1)}$ if and only if $z^{*(k+1)} = \overline{z^{(k+1)}}$. A fault propagation happens only if the fault is ``active''
\end{itemize}
}

\textcolor{black}{
\textcolor{black}{The main operation of $\texttt{A2B}$ is to compute $z = x + y \bmod{2^{(k+1)}}$, and it performs this by utilizing Boolean masking. However, for our purpose, we just need to focus on the addition functionality and the arithmetic shares ($x$ and $y$).} We can write: 
\begin{equation} \label{eq1}
    z^{(j)} = x^{(j)} \xor y^{(j)} \xor c^{(j-1)},
\end{equation}
where $1\leq j\leq (k+1)$, $c^{(j-1)}$ is the carry for $(j-1)$-th bit, and $c^{(0)} = 0$.
The carry can be written further as: 
\begin{equation} \label{eq2}
\begin{split}
c^{(j-1)} &= (x^{(j-1)} \& y^{(j-1)}) \xor ((x^{(j-1)} \xor y^{(j-1)}) \& c^{(j-2)}) \\ 
&= (x^{(j-1)} \& y^{(j-1)}) \xor (x^{(j-1)} \& c^{(j-2)}) \xor (y^{(j-1)} \& c^{(j-2)})\\
&= ((y^{(j-1)} \xor c^{(j-2)}) \& x^{(j-1)}) \xor (y^{(j-1)} \& c^{(j-2)})\,.
\end{split}
\end{equation}
The relation between $x,\ y,\ z$ and $m_1,\ m_2,\ \hat{m}$ is given as:
\begin{equation} \label{eq3}
    z^{(k+1)} = x^{(k+1)} \xor y^{(k+1)} \xor c^{(k)} = m_1 \xor m_2 = \hat{m}\,,
\end{equation}
\begin{equation} \label{eq3.5}
    c^{(k)} = ((y^{(k)} \xor c^{(k-1)}) \& x^{(k)}) \xor (y^{(k)} \& c^{(k-1)})\,.
\end{equation}
}

\textcolor{black}{Next, we present the following lemmas:}

\textcolor{black}{
\begin{lemma} \label{lemma:stuck-at-1-for-k}
If we introduce a \texttt{stuck-at-1} fault at $x^{(k-1)}$, the fault activates with probability $\frac{1}{2}$. The activated fault propagates to $z^{(k)}$ (\textit{i.e.} $z^{*(k)} = \overline{z^{(k)}}$) if and only if $z^{(k-1)} = 1$. 
\end{lemma}}
\begin{proof}
\textcolor{black}{The introduced \texttt{stuck-at-1} fault at $x^{(k-1)}$ is \emph{active}, \textit{i.e.} $x^{*(k-1)} = \overline{x^{(k-1)}}$, only if $x^{(k-1)}$ is equal to $0$. \textcolor{black}{Since $x$ is a random arithmetic share, $x^{(k-1)} = 0$~happens with probability $\frac{1}{2}$.} If the fault is active then $x^{*(k-1)} = 1$. From Equation~\ref{eq1}, we get that the introduced fault at $x^{(k-1)}$ can only propagate to $z^{(k)}$ through $c^{(k-1)}$. Let us assume $c$ becomes $c^*$ after the fault injection. Now, from Equation~\ref{eq2}, we get $c^{*(k-1)} = ((y^{(k-1)} \xor c^{(k-2)}) \& x^{*(k-1)}) \xor (y^{(k-1)} \& c^{(k-2)})$. The fault propagation (i.e. $c^{*(k-1)} = \overline{c^{(k-1)}}$) happens only if $(y^{(k-1)} \xor c^{(k-2)}) = 1$. This implies that one of $y^{(k-1)}$ and $c^{(k-2)}$ is zero, \textit{i.e.} $(y^{(k-1)} \& c^{(k-2)}) = 0$. So, from Equation~\ref{eq1}, we get that $z^{(k-1)} = x^{(k-1)} \xor (y^{(k-1)} \xor c^{(k-2)}) = 0 \xor 1 = 1$. \textcolor{black}{This is the unmasked value of the $(k-1)$th bit of the coefficient that we target. The fault injection undoes the masking for this specific bit, thanks to the fault propagation. Overall, the fault reaches to $z^{(k)}$ with probability $\frac{1}{2}$ when $z^{(k-1)} = 1$}.}
\end{proof}

\textcolor{black}{
\begin{lemma} \label{lemma:stuck-at-1-for-k+1}
If we introduce a \texttt{stuck-at-1} fault at $x^{(k-1)}$, the fault activates with probability $\frac{1}{2}$. The activated fault propagates to $z^{(k+1)}$, \textit{i.e.} ($z^{*(k+1)} = \overline{z^{(k+1)}}$) if and only if $z^{(k-1)} = 1$ and $z^{(k)} = 1$. 
\end{lemma}}
\begin{proof}
\textcolor{black}{
\textcolor{black}{According to Lemma~\ref{lemma:stuck-at-1-for-k}, the fault at $x^{(k-1)}$ is active with probability $\frac{1}{2}$. Also, the activated fault propagates only if $z^{(k-1)} = 1$. Now, $z^{(k+1)} = x^{(k+1)} \xor y^{(k+1)} \xor c^{(k)}$. The fault can only reach $z^{(k+1)}$ through $c^{(k)}$.  Next we observe that $c^{(k)} = ((y^{(k)} \xor c^{(k-1)}) \& x^{(k)}) \xor (y^{(k)} \& c^{(k-1)})$. Therefore, the only path for the fault to reach $c^{{k}}$ from $x^{(k-1)}$ is via $c^{(k-1)}$. Now as shown in Lemma~\ref{lemma:stuck-at-1-for-k}, $c^{(k-1)}$ gets corrupted (i.e. $c^{(k-1)} \neq c^{*(k-1)}$) only if $x^{(k-1)} = 0$. So we only need to determine when a fault in $c^{(k-1)}$ propagates to $c^{(k)}$. According to the equation of $c^{(k)}$ (Equation~\ref{eq3.5}), this is only possible if $x^{(k)} \xor y^{(k)} = 1$. Finally, we observe that $c^{(k-1)} = ((y^{(k-1)} \xor c^{(k-2)}) \& x^{(k-1)}) \xor (y^{(k-1)} \& c^{(k-2)}) = (1 \& 0) \xor 0 = 0$ (from the proof of Lemma~\ref{lemma:stuck-at-1-for-k}). Therefore, it is evident that $z^{(k)} = x^{(k)} \xor y^{(k)} \xor c^{(k-1)} = 1$. This proves the lemma.
%
}
}
%
%
%
%
\end{proof}

\noindent\textcolor{black}{\textbf{Example:} Now, let us provide a very simple example to visualize Lemma~\ref{lemma:stuck-at-1-for-k+1}. Let us consider $k = 2$ (therefore, we have a 3-bit register). In this case, we introduced a \texttt{stuck-at-1} fault at $x^{(1)}$. The fault is active only if $x^{(1)}$ is equal to $0$. If the fault is active then $z^* = (z + 1) \bmod{2^3}$. Table~\ref{tab:value-z-after-fault} presents the values of $z$ and $z^{(3)}$ before the \texttt{stuck-at-1} fault at $x^{(1)}$ together with the values of $z^*$ and $z^{*(3)}$ after the fault, when the \texttt{stuck-at-1} fault at $x^{(1)}$ is active. From this example, we can observe that the fault at $x^{(1)}$ only propagates to $z^{(3)}$, if $z^{(1)} = 1$ and $z^{(2)} = 1$.} 

\begin{table}[httb]
\centering
\caption{The value of $z$ and $z^{(3)}$ before and after the \texttt{stuck-at-1} fault at $x^{(1)}$}
\label{tab:value-z-after-fault}
\begin{tabular}{|cccc|ccccc|}
\hline
\multicolumn{4}{|c|}{Before the fault injection}                                                                                                 & \multicolumn{5}{c|}{After the fault injection}                                    \\ \hline
\multicolumn{1}{|c|}{$z^{(3)}$}           & \multicolumn{1}{c|}{$z^{(2)}$}           & \multicolumn{1}{c|}{$z^{(1)}$} & $z$                      &  &  & $z^{*}$                  & \multicolumn{1}{c|}{} & $z^{*(3)}$               \\ \hline
\multicolumn{1}{|c|}{}                    & \multicolumn{1}{c|}{}                    & \multicolumn{1}{c|}{0}         & 0                        &  &  & 1                        & \multicolumn{1}{c|}{} & 0                        \\ \cline{3-9} 
\multicolumn{1}{|c|}{}                    & \multicolumn{1}{c|}{\multirow{-2}{*}{0}} & \multicolumn{1}{c|}{1}         & 1                        &  &  & 2                        & \multicolumn{1}{c|}{} & 0                        \\ \cline{2-9} 
\multicolumn{1}{|c|}{}                    & \multicolumn{1}{c|}{}                    & \multicolumn{1}{c|}{0}         & {\color[HTML]{333333} 2} &  &  & 3                        & \multicolumn{1}{c|}{} & 0                        \\ \cline{3-9} 
\multicolumn{1}{|c|}{\multirow{-4}{*}{0}} & \multicolumn{1}{c|}{\multirow{-2}{*}{1}} & \multicolumn{1}{c|}{1}         & {\color[HTML]{333333} 3} &  &  & {\color[HTML]{CB0000} 4} & \multicolumn{1}{c|}{} & {\color[HTML]{CB0000} 1} \\ \hline
\multicolumn{1}{|c|}{}                    & \multicolumn{1}{c|}{}                    & \multicolumn{1}{c|}{0}         & 4                        &  &  & 5                        & \multicolumn{1}{c|}{} & 1                        \\ \cline{3-9} 
\multicolumn{1}{|c|}{}                    & \multicolumn{1}{c|}{\multirow{-2}{*}{0}} & \multicolumn{1}{c|}{1}         & 5                        &  &  & 6                        & \multicolumn{1}{c|}{} & 1                        \\ \cline{2-9} 
\multicolumn{1}{|c|}{}                    & \multicolumn{1}{c|}{}                    & \multicolumn{1}{c|}{0}         & {\color[HTML]{333333} 6} &  &  & 7                        & \multicolumn{1}{c|}{} & 1                        \\ \cline{3-9} 
\multicolumn{1}{|c|}{\multirow{-4}{*}{1}} & \multicolumn{1}{c|}{\multirow{-2}{*}{1}} & \multicolumn{1}{c|}{1}         & {\color[HTML]{333333} 7} &  &  & {\color[HTML]{CB0000} 0} & \multicolumn{1}{c|}{} & {\color[HTML]{CB0000} 0} \\ \hline
\end{tabular}
\end{table}

\textcolor{black}{
\begin{lemma} \label{lemma:stuck-at-1-for-k+1-msg}
A \texttt{stuck-at-1} fault at $x^{(k-1)}$ activates with probability $\frac{1}{2}$ and propagates to $z^{(k+1)}$ and causes decryption failure if and only if $z^{(k-1)} = 1$ and corresponding message bit $\hat{m} = 1$. 
\end{lemma}}
\begin{proof}
\textcolor{black}{
From Equation~\ref{eq:k=k+1} and Equation~\ref{eq3}, we have $z^{(k)} = z^{(k+1)} = \hat{m}$. So, the \texttt{stuck-at-1} fault at $x^{(k-1)}$ propagates to $z^{(k+1)} = \hat{m}$, only if $z^{(k-1)} = 1$ and the corresponding message bit is $1$ (according to the previous lemmas). As the fault propagates to $\hat{m}$ then $\hat{m^*} = \overline{\hat{m}}$. This event will cause decryption failure as the decoded message bit is different before and after the fault injection. \textcolor{black}{Once again we remind that $\hat{m}$ will remain as two shares. However, by the correctness of the Boolean masking, the properties described in these lemmas will follow.}}
\end{proof}

\textcolor{black}{
In a nutshell, fault propagation provides information about $z^{(k-1)}$, and propagates till $z^{(k+1)}$ conditioned on $z^{(k)}$. Further, since $z^{(k)}$ and $z^{(k+1)}$ are the message bits (and assumed known in our attacks), if corrupted, they lead to decryption failure. Overall, we obtain a (unmasked) bit ($z^{(k-1)}$) which exposes partial information about the secret-dependent noise. Notably, we receive similar results if we use \texttt{stuck-at-0} fault. However, the fault propagation would happen when $z^{(k-1)} = 0$, in that case.} 
%
%
%
%
\textcolor{black}{However, bit-flip faults do not provide information in this case. For a \texttt{bit-flip} fault at $x^{(k-1)}$, it is active for both possible cases, when $x^{(k-1)} = 0$ or when $x^{(k-1)} = 1$. The injected fault will propagate to $z^{(k)}$ and cause successful decryption failure for $2^{k-1}$ instances out of $2^k$ instances when $z^{(k-1)} = 1$ and also when $z^{(k-1)} = 0$. Hence, successful decryption failure does not provide knowledge regarding $z^{(k-1)}$.} 

\subsection{Fault Propagation for the Higher-Order Masking}

\textcolor{black}{We now discuss the effect of our \texttt{stuck-at} fault on the higher-order masked decapsulation algorithms. In $t$-th order masking, the input of \texttt{A2B} conversion $g[i]$ has $t+1$ arithmetic shares. For the sake of simplicity, let us assume $z = g[i]$ and its $(t+1)$ arithmetic shares are $(x_1,\ x_2,\ \ldots,\ x_{t+1})$. So, $(\sum_{j=1}^{(t+1)} x_j ) \bmod{2^{(k+1)}} = z$. Now, this multi-share sum can be written as a two-share case $z = (x_1 + y) \bmod{2^{(k+1)}}$, where $y:= (\sum_{j=2}^{(t+1)} x_j ) \bmod{2^{(k+1)}}$. This brings back the two-share scenario explained in the previous subsection. More precisely, we introduce a \texttt{stuck-at-1} fault at $x_1^{k-1}$. The fault is active only if $x_1^{k-1} = 0$. The injected fault at $x_1^{k-1}$ will not affect $y$, and can propagate to $z^{(k+1)}$ only through $x_1^{k-1}$.
%
%
%
Therefore, by applying Lemma~\ref{lemma:stuck-at-1-for-k+1-msg}, the \texttt{stuck-at-1} fault at $x_1^{k-1}$ will propagate to $z^{k+1}$ only if $z^{(k-1)} = 1$ and $z^{(k)} = 1$. The conclusion is that the fault propagation scenario we describe here does not get affected by the order of masking.}

\subsection{Application of Our Fault Attack on Kyber}\label{subsec:fault-attack-kyber}

\textcolor{black}{
In Kyber, $q = 3329$ and hence $k = 12$. So, each coefficient of $m'\in \mathbb{R}_q$ are of $12$ bits. \textcolor{black}{However, in order to enable message extraction in a masked setting, the shares are transferred from mod $q$ to mod $2^{(k+1)}$, making each share a 13-bit value.} \emph{We inject \texttt{stuck-at-1} bit-fault at the $11$th bit of the $i$-th coefficient of an input share of the \texttt{A2B} of the masked \texttt{Decode} algorithm.} Whether or not the fault propagates to the 13-th bit (which remains in a Boolean shared form) exposes the 11-th bit. 
%
}

\begin{figure}[!tbp]
  \centering
  \subfloat[\texttt{stuck-at-1} fault]{\includegraphics[width=0.4\textwidth]{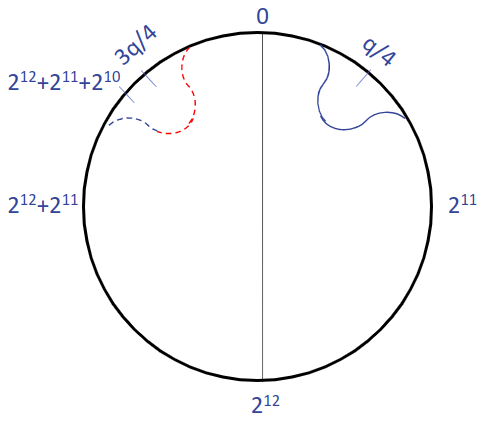}\label{fig:kyber-fault-at1-2}}
  \hfill
  \subfloat[\texttt{stuck-at-0} fault]{\includegraphics[width=0.4\textwidth]{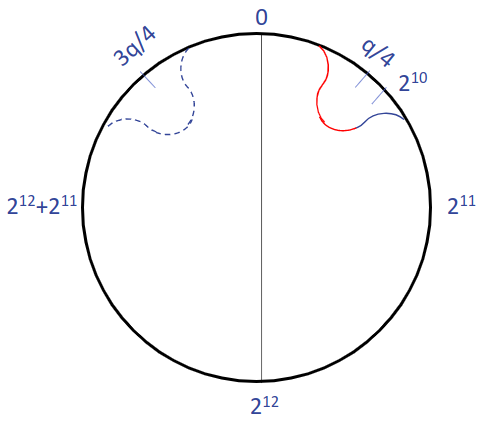}\label{fig:kyber-fault-at0-2}}
  \caption{Results of \texttt{stuck-at} fault attacks on Kyber. The solid curve represents the probability distribution of a coefficient that decoded to the message value $m[i]=0$ and the dashed curve represents the probability distribution of a coefficient that decoded to the message value $m[i]=1$. This picture also presents that the injected fault will corrupt the message value whenever it lies on the red part.}
\end{figure}

\textcolor{black}{
According to our nomenclature defined in the previous section, we denote the target 11-th bit (to be faulted) as $g_{a_1}[i]^{(11)}$, and the 11-th-bit value to be extracted as $g[i]^{(11)}$, where $g = (g_{a_1} + g_{a_2}) \bmod{2^{13}}$.
From Lemma~\ref{lemma:stuck-at-1-for-k+1-msg}, we obtain that the \texttt{stuck-at-1} fault at $g_{a_1}[i]^{(11)}$ propagates to the corresponding $i$-th message bit $m[i]$ and causes decryption failure ( with probability $1/2$) when $g[i]^{(11)} = 1$ and $m[i] = 1$. Figure \ref{fig:kyber-fault-at1-2} presents the value ranges for $g[i]$, where the \texttt{stuck-at-1} fault will propagate and cause decryption failure, and in which cases the fault will not cause decryption failure for Kyber. The value range of $g[i]$, which will cause decryption failure are $\{2^{12}+2^{11}+2^{10},\  2^{12}+2^{11}+2^{10}+1,\ \ldots,\ 2^{13}-1\}$ (coloured in red). The value range of $g[i]$, which will not cause decryption failure are $\{2^{13}-\lceil q/2 \rceil,\  2^{13}-\lceil q/2 \rceil+1,\ \ldots,\ 2^{12}+2^{11}+2^{10}-1\}$(coloured in blue).} \textcolor{black}{One important point is that the distributions are bell-shaped, i.e. the mean values occur more often than any other value, and the values at the tails are extremely rare. This fact will be used later for relaxing the fault model to tolerate multi-bit faults.} \textcolor{black}{Similarly, the \texttt{stuck-at-0} fault at $g_{a_1}[i]^{(11)}$ propagates to $m[i]$ and causes decryption failure when $g[i]^{(11)} = 0$ and $m[i] = 0$ with probability $1/2$. Figure \ref{fig:kyber-fault-at0-2} shows the value of $g[i]$ where the \texttt{stuck-at-0} fault will propagate and cause decryption failure and where the fault will not cause decryption failure. The values of $g[i]$, which will cause decryption failure are $\{0,\ 1,\ \ldots,\ 2^{10}-1\}$ (coloured in red) and only when $m[i]=0$ and the fault is active.}
\textcolor{black}{There are some implementations of masked Kyber where $16$ bit register is used for $g[i]$ instead of $13$ bit. One such implementation is~\cite{FO-masked-kyber-HeinzKLPSS22} (We used this implementation in our practical setup). We note that if the \texttt{stuck-at-1} fault at $g[i]^{(11)}$ propagates to $g[i]^{(13)}$ then it will propagate to $g[i]^{(16)}$. This can be proven by repeated application of Lemma~\ref{lemma:stuck-at-1-for-k+1}. Therefore, working with the 13th or the 16th bit is the same for us, and we stick with the 13th bit case.}

\textcolor{black}{
Now, we recall that the 11th bit corresponds to the (secret dependent) noise. \textcolor{black}{\emph{After the injection of the \texttt{stuck-at-1} fault at $g[i]^{(11)}$, it will cause a decryption failure if $g[i]^{(11)} = 1$ and the corresponding message bit $m[i] = 1$. If $m[i] = 1$ and $g[i]^{(11)} = 1$, then the associated decryption noise is $\geq \lfloor q/4 \rfloor- ( 2^{10} - \lfloor q/4 \rfloor)$.}} So, it can create a partition in the distribution of the decryption noise depending on the decryption failure. This partition leaks some information regarding the secret key, \textcolor{black}{and we can eventually form inequalities involving the secret key (described in the next subsection).} Note that the partition we received from our fault attack is different from the attacks available in the literature~\cite{PesslP21,HermelinkPP21,Delvaux22}. In the previous works, if the message bit $m[i] = 1$, then after the injection of the fault for almost half of the possible values of $g[i]$ (around $q/4$ values out of $q/2$ values) the fault will cause decryption failure, and for the other half of the possible values of $g[i]$ will not cause decryption failure. The same incident happens when the message bit $m[i] = 0$. However, in our \texttt{stuck-at-1} attack, we observed decryption failures only when the corresponding message bit $m[i] = 1$. If $m[i] = 1$, then in our experiments, we have noticed $99.3\%$ decryption failure and only $0.7\%$ decryption success by applying \texttt{stuck-at-1} fault at $11$th bit of a share of $g[i]$ for Kyber512. We provide more details in Section~\ref{sec:experiments}.} 

\subsection{Recovery of the Secret Key for Kyber}\label{sec:inequality-solve}

\textcolor{black}{This subsection will describe the final stage of our attack \textit{i.e.} recovering the secret key for Kyber. As mentioned earlier, decryption noise forms a linear equation with the secret key $\pmb{s}$ and error $\pmb{e}$. The decryption noise for LWE-based KEM is equal to $\pmb{e}^T \pmb{s}'[i] - \pmb{s}^T(\pmb{e_1}[i] + \Delta \pmb{u}[i]) + e_2[i] + \Delta v[i]$, where $\Delta \pmb{u}$ and $\Delta {v}$ are the noise introduced because of the compression operation over $\pmb{u}$ and $v$, respectively. In our case, a decryption failure is observed only when the message bit $m[i] = 1$, and the corresponding 11-th bit is also 1. 
If the fault is active and decryption failure happens, then the decryption noise satisfies the following inequality.
\begin{equation}
    (\pmb{e},\ \pmb{s})^T (\pmb{s}'[i],\  - (\pmb{e_1}[i] + \Delta \pmb{u}[i])) + e_2[i] + \Delta v[i] \geq \lfloor q/4 \rfloor - ( 2^{10} - \lfloor q/4 \floor). \label{eq:ky1}
\end{equation}
}
\textcolor{black}{
If the fault is active and decryption success happens, then the decryption noise satisfies the following inequality.
\begin{equation}
    (\pmb{e},\ \pmb{s})^T (\pmb{s}'[i],\  - (\pmb{e_1}[i] + \Delta \pmb{u}[i])) + e_2[i] + \Delta v[i] < \lfloor q/4 \rfloor - ( 2^{10} - \lfloor q/4 \rfloor). \label{eq:ky2}
\end{equation}
Note that the fault is active with probability $\frac{1}{2}$.
\textcolor{black}{As mentioned in Section \ref{sec:intro},} the attacker has access to the encapsulation oracle and works with self-generated ciphertexts. So, the attacker knows all the values in Equation \ref{eq:ky1} \& \ref{eq:ky2}, except $(\pmb{e},\ \pmb{s})$. The attacker can find the polynomials $(\pmb{e},\ \pmb{s})$ by solving these linear inequalities.}

\textcolor{black}{Recently, a few inequality solvers have been proposed~\cite{PesslP21,HermelinkPP21,Delvaux22,HermelinkMSPR23}. All these solvers are based on BP in order to tolerate errors and solve the system with a large number of inequalities in a reasonable amount of time. In this work, we have used the inequality solver proposed by Delvaux et al. in \cite{Delvaux22} to solve the inequalities obtained from the \texttt{stuck-at-1} fault on Kyber.
We always inject the fault at some fixed coefficient. 
If we consider $\omega$ numbers of inequalities, then the system of inequality contains $\omega$ inequalities with $\psi$ unknowns. Here the value of $\psi$ is the total number of coefficients of $\pmb{s}$ and $\pmb{e}$, which is $2*l*256$, $l=2$ for Kyber-$512$. This inequality solver uses an iterative method to solve the system of $\omega$ inequalities by maintaining a  probability mass function (PMF) for each of $\psi$ unknowns. The PMF for each unknown initiated with the CBD on $[-\eta,\ \eta]$ and updated in each iteration. The updation of PMF for each unknown is performed by calculating the probability to satisfy each of the $\omega$ inequalities for each possible value $[-\eta,\ \eta]$ of the unknown, and then all the $2*\eta*\omega$ probabilities are combined. \cite{Delvaux22} accelerates the computation time of their solver by applying the central limit theorem. The iteration of the updation of PMF for each unknown continues until a suﬃciently approximate one-point distribution on $[-\eta,\ \eta]$ is found.}

\subsection{Application of Our \texttt{stuck-at} Fault Attack on Saber}

\begin{figure}[!tbp]
  \centering
  \subfloat[Stuck-at-1 fault]{\includegraphics[width=0.35\textwidth]{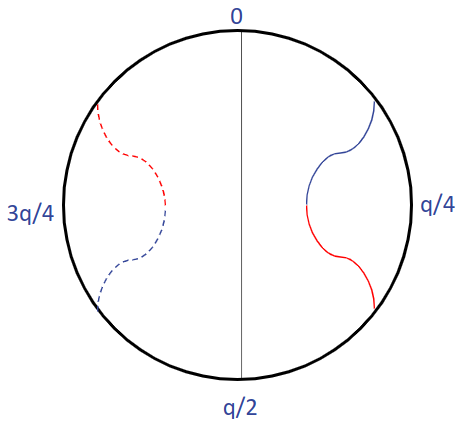}\label{fig:stuck-at-1}}
  \hfill
  \subfloat[Stuck-at-0 fault]{\includegraphics[width=0.35\textwidth]{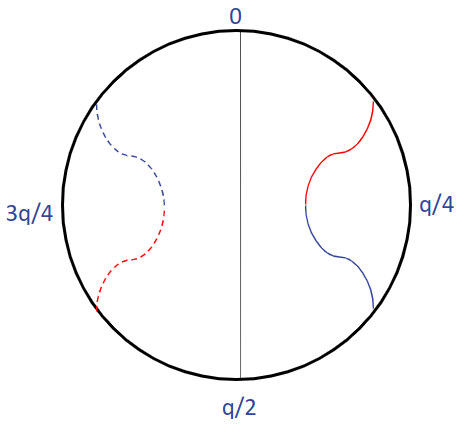}\label{fig:stuck-at-0}}
  \caption{Results of \texttt{stuck-at} fault attacks on Saber. The probability distribution of $m'[i]$ is shown in these figures. The solid curve represents all the values of $m'[i]$ that decoded to the message value $0$, and the dashed curve represents all the values of $m'[i]$ that decoded to the message value $1$. The inserted fault will corrupt the message value whenever the value of $m'[i]$ lies on the red part.}
\end{figure}

\textcolor{black}{In Saber, $q = p = 2^{\epsilon_p}$ is a power-of-two moduli and $k = \epsilon_p$ ($\epsilon_p = 10$ for the medium security version of Saber). The decoding algorithm in Saber has only two steps. The input of the decoding algorithm of Saber is arithmetic shares of $m' \in \mathbb{R}_p$. First, the algorithm performs the \texttt{A2B} conversion on the arithmetic shares of $m'$ to convert them into Boolean shares. Then the most significant bit ($\epsilon_p$th bit) is extracted from the Boolean shares of $m'$ by using $(\epsilon_p-1)$ right shift operation. This final output is the message polynomial $m$, which is the output of the decoding in Saber. $m'[i]$ follows a bell-shaped distribution with a peak at $p/4$ when $m'[i]^{(\epsilon_p)}$ is 0, \textit{i.e.} for the corresponding message bit $0$. The values of $m'[i]$ follow a bell-shaped distribution with a peak at $3p/4$ when $m'[i]^{(\epsilon_p)}$ is 1, \textit{i.e.} for the corresponding message bit $1$. From Lemma~\ref{lemma:stuck-at-1-for-k}, we get that the injected \texttt{stuck-at-1} fault at the $(\epsilon_p-1)$-th bit ($9$th bit for the medium security version of Saber) of a share of $m'[i]$ propagates and causes decryption failure if it is active and $m'[i] \in \{(p/4),\ (p/4)+1,\ \ldots,\ (p/2)-1\} \cup \{3(p/4),\ 3(p/4)+1,\ \ldots,\ p-1\}$ (coloured in red). The injected fault is active with $1/2$ probability. Also, if $m'[i] \in \{0,\ 1,\ \ldots,\ (p/4)-1\} \cup \{(p/2),\ (p/2)+1,\ \ldots,\ 3(p/4)-1\}$, then the fault does not propagate and hence does not cause decryption failure (coloured in blue). We use Figure \ref{fig:stuck-at-1} to show the values of $m'[i]$ in which the \texttt{stuck-at-1} fault propagation is successful and the values of $m'[i]$ in which this fault will not cause decryption failure. In this figure, we use a solid bell curve to denote the distribution for the message bit $0$ and a dashed bell curve to indicate the distribution for the message bit $1$. Similarly, we get that the injected \texttt{stuck-at-0} fault at the $(\epsilon_p-1)$-th bit of a share of $m'[i]$ propagates and causes decryption failure if it is active and $m'[i] \in \{0,\ 1,\ \ldots,\ (p/4)-1\} \cup \{(p/2),\ (p/2)+1,\ \ldots,\ 3(p/4)-1\}$ (coloured in red). Also, if $m'[i] \in \{(p/4),\ (p/4)+1,\ \ldots,\ (p/2)-1\} \cup \{3(p/4),\ 3(p/4)+1,\ \ldots,\ p-1\}$ then decryption failure does not arise(coloured in blue). Figure \ref{fig:stuck-at-0} represents the values of $m'[i]$ for both cases when the \texttt{stuck-at-0} fault propagation is successful and when fault propagation is not. \textcolor{black}{Overall, we see that the fault propagation and the nature of the information leakage are quite similar to that of Kyber's.}
This gives enough evidence that our fault attack can be applied to any LWE-based KEM with some minor changes, as the leakage pattern from the masked decode algorithm is similar for all these schemes, such as we can use Kyber's analysis to NewHope~\cite{newhope} just by changing the parameters. However, we left this for the extended version of this work.   
\textcolor{black}{Hereon, we shall mainly focus on the Kyber scenario in this paper and develop the BP-based key recovery algorithm for it.}}

%% file: images/masked_scheme.tex
\begin{figure}[!ht]
    \centering
\setlength{\nodedistance}{4mm}
\resizebox{0.75\textwidth}{!}{
\begin{tikzpicture}[ 
node distance = \nodedistance,
operation/.style = {circle, draw=black},
bigoperation/.style = {rounded corners, draw=black},
operationsecret/.style = {circle, draw=black, fill=black!20!white, text=black, very thick},
bigoperationsecret/.style = {rounded corners, draw=black, fill=black!20!white, text=black, very thick},
Bigoperationsecret/.style = {rounded corners, draw=black, fill=black!20!white, text=black, very thick, minimum height=4\nodedistance, , minimum width=2\nodedistance},
operant/.style = {},
halfblockdraw/.style = {draw, rounded corners},
line/.style = {draw, -Latex},
connect with angle/.style=
{to path={let \p1=(\tikztostart), 
\p2=(\tikztotarget) in -- ++(0, {(\x2-\x1)*tan(#1-90)-(\y1-\y2)}) -- (\tikztotarget)}},
connect with angle hor/.style=
{to path={let \p1=(\tikztostart), 
\p2=(\tikztotarget) in -- ++({(\x2-\x1)-(\y1-\y2)*tan(#1-90)}, 0) -- (\tikztotarget)}},
                       ]

\node (decom) [bigoperation]  {$\mathtt{Decomp}$};
\node (bp) [operant, left=of decom.west]  {$\pmb{u}$};
\draw[->] (bp) -- (decom);

\node (multbps) [operationsecret, right=of decom]  {X};
\draw[->] (decom) -- (multbps);


\node (s1) [operant, above=of multbps.north]  {$\pmb{s}_1$};
\draw[->, very thick] (s1) -- (multbps);



\node (multbps2) [operationsecret, below=of multbps]  {X};
\draw[->] (decom) -- (decom |- multbps2) -- (multbps2);

\node (s2) [operant, below=of multbps2]  {$\pmb{s}_2$};
\draw[->, very thick] (s2) -- (multbps2);

\coordinate  (addvh) at ($(multbps.east)$);


\node (cm) [operant, below=5\nodedistance of bp]  {$v$};
\node (shiftcm) [bigoperation, right=of cm]  {$\mathtt{Decomp}$};
\draw[->] (cm) -- (shiftcm);

\coordinate [right=of addvh](addvhright);
\node (subcmv) [operationsecret, right=of multbps2] {$-$};
\draw[->, very thick] (multbps2) -- (subcmv);
\draw[->] (shiftcm) -- (subcmv |- shiftcm) -- (subcmv);

\node (tobin) [Bigoperationsecret, right=2\nodedistance of $(subcmv.east)!0.5!(addvh.east)$]  {$\mathtt{Decode}$};
\draw[->, very thick] (addvh) -- (tobin.west |- addvh);
\draw[->, very thick] (subcmv) -- (tobin.west |- subcmv);

\node[draw=red, -, very thick, fit=(tobin)] {};

\node (g) [Bigoperationsecret, right= 2\nodedistance of tobin]  {$\mathcal{G}$};
\node (pkh) [operant, above=of g]  {$pkh$};
\draw[->, transform canvas={xshift=0.5em}] (pkh) -- (g);
\draw[->, transform canvas={xshift=-0.5em}] (pkh) -- (g);
\draw[->, very thick] (tobin.east |- addvh) -- (g.west |- addvh);
\draw[->, very thick] (tobin.east |- subcmv) -- (g.west |- subcmv);

\coordinate  (m1) at ($(tobin.east |- addvh)!1.0!(g.west |- addvh)$);
\coordinate  (m2) at ($(tobin.east |- subcmv)!0.4!(g.west |- subcmv)$);

\node (K) [operant, below=of g]  {$\hat{K}'$};
\draw[->, very thick, transform canvas={xshift=0.5em}] (g) -- (K);
\draw[->, very thick, transform canvas={xshift=-0.5em}] (g) -- (K);

\node (reenc) [Bigoperationsecret, right=of g]  {$\mathtt{Re-encryption}$};
\draw[->, very thick] (g.east |- addvh) -- (reenc.west |- addvh);
\draw[->, very thick] (g.east |- subcmv) -- (reenc.west |- subcmv);

\coordinate [below=0.6\nodedistance of K](tmp2);
\coordinate [below=0.2\nodedistance of K](tmp1); 
\coordinate [below=2.6\nodedistance of reenc](tmp3); 
\draw[->, very thick, transform canvas={xshift=-1.0em}] (tmp3) -- (reenc);
\draw[-, very thick, transform canvas={xshift=-1.0em}] (m1) -- (m1 |- tmp1);
\draw[-, very thick, transform canvas={xshift=-1.0em}] (m1 |- tmp1) -- (tmp3);
\draw[->, very thick] (m2) -- (m2 |- tmp2) -- (reenc |- tmp2) -- (reenc);

\coordinate[right= 2\nodedistance of reenc] (output);
\node (bp2) [operant, right=of $(reenc.east |- addvh)$] {$(\pmb{u}_{*1},\pmb{u}_{*2})$};
\node (cm2) [operant, right=of $(reenc.east |- subcmv)$]  {$({v}_{*1},{v}_{*2})$};
\draw[->, very thick] (reenc.east |- bp2) -- (bp2);
\draw[->, very thick] (reenc.east |- cm2) -- (cm2);

\node[draw=blue, dashed, thick, fit=(s1) (shiftcm) (tobin)](ciphertext) {};

\node[draw=black, dotted, thick, fit=(bp2) (cm2)](ciphertext2) {};
\node[draw=black, dotted, thick, fit=(bp) (cm)](ciphertext) {};
\coordinate [below=2\nodedistance of K](x); 
\coordinate (y) at ($(ciphertext)!0.5!(ciphertext2)$);
\node (comp) [operationsecret, very thick] at (y |- x)  {$=?$};

\draw[->] (ciphertext) -- (ciphertext |- x) -- (comp);
\draw[->, very thick] (ciphertext2) -- (ciphertext2 |- x) -- (comp);

\node (correct) [bigoperation, below left=2\nodedistance and 2\nodedistance of comp]  {return $\mathcal{H}(\hat{K}', c)$};
\draw[->] (comp.south) -- (correct);
\node[operant,  below left=0.5\nodedistance and 1.3\nodedistance of comp] (yes) {yes};

\node (correct) [bigoperation, below right=2\nodedistance and 2\nodedistance of comp]  {return $\mathcal{H}(z, c)$};
\draw[->] (comp.south) -- (correct);
\node[operant,  below right=0.5\nodedistance and 1.3\nodedistance of comp] (no) {no};
\end{tikzpicture}}

\caption{First-order masked decapsulation algorithm of LW(E/R) based schemes. The highlighted operations in color gray are influenced by the non-ephemeral secret-key $\pmb{s} = (\pmb{s}_1,\ \pmb{s}_2)$. The function we exploit in our attack is enveloped with the red rectangle.}

\label{fig:maskeddecaps}
\end{figure}

%% file: sections/relax_fa.tex
\vspace{-10pt}
\section{Relaxation of the Fault Model}\label{sec:relaxed-fault}
\textcolor{black}{{The attack described so far in this paper requires faults to be injected at a specific bit (the $11$th bit) of a register. While such a precise injection has been demonstrated previously in literature~\cite{dutertre2018laser,breier2022practical,saha2020fault}, injection at a specific bit is always probabilistic and depends on the precision of the injection mechanism}. Therefore, we can only expect the fault to happen at the $11$th bit with some probability $p$. On the other hand, research has shown that limiting the width of faults to one bit is feasible~\cite{ghalaty2014differential} even with low-cost injection mechanisms such as clock-glitch. However, the location of the faulted bits is not well-controllable in many situations. \textcolor{black}{Furthermore, with low-cost setups, multi-bit faults also occur quite frequently.} Therefore, it is practical to assume a fault model that injects single-bit faults at random locations, or \textcolor{black}{even multi-bit} in a register. \textcolor{black}{Depending on the injection mechanism, the fault model may vary. However, \texttt{stuck-at} faults can be observed in several practical scenarios, such as clock-glitch or EM. Even in our practical experiments with EM injection described in Section~\ref{sec:experiments}, we observed single/multi-bit \texttt{stuck-at-1} faults.}}


\textcolor{black}{We now analyze the effect of such a relaxed fault model (single/multi-bit faults at different locations) assumption on the proposed attack. One should note that the proposed fault attack utilizes the inequalities constructed from the decryption failures/successes. There is no direct way to distinguish whether or not the fault has been injected at the desired location ($11$th bit for Kyber) in the case of a decryption failure. Similarly, in the case of a decryption success, it is hard to determine if it is due to the fault at $11$-bit or some other fault. The injections at the undesired locations may generate \emph{noisy} equations, hindering the convergence of the BP algorithm to the correct key value. \textcolor{black}{Generally, BP algorithms can tolerate noise only up to a certain extent~\cite{HermelinkPP21,Delvaux22}.} }

\textcolor{black}{Interestingly, we have observed that \textcolor{black}{our attack is not significantly affected by such injection noise. The \emph{signal}, in our case, is a single-bit injection (\texttt{stuck-at-1}) at the 11th bit.} The main reason behind such observation is that due to the carry chain structure, many faults do not propagate till the most significant bit of the addition operation. More precisely, the only way to observe a fault is to observe it at the MSB bit, and a fault in some least significant bits can propagate to the MSB only through the carry chain. The carry chain contains several AND operations, which makes the fault propagation conditional on many bits, making it rare. It is also due to the normally distributed data that we are dealing with. The target variable assumes only a specific set of values with high probability among the entire range, which further hinders the fault propagation from most of the least significant bits.
The probability of fault propagation is the highest for the two most significant bits, $12$th and $13$th bit for Kyber, and the second highest is for the next significant bit, $11$th bit for Kyber. However, for the lower-order bits, the probability of fault propagation is negligible. In the next few paragraphs, we further elaborate on why the fault propagation probability is low for lower-order bits.}

\textcolor{black}{ 
\textcolor{black}{For the time being, let us further narrow down our discussion to Kyber-512, although the conclusions made would be similar for other variants.}
%
\textcolor{black}{As already mentioned in Section~\ref{sec:inequality-solve}, we extract a single-bit of the decryption noise via faults, which provides us with some information on the secret key.} 
For Kyber512, the mean of the decryption noise is $0$ associated with any of the message bits. Furthermore, decryption noise  remains between $(-\lfloor \frac{q}{4} \rfloor,\ \lfloor \frac{q}{4} \rfloor)$ with probability $(1-\frac{1}{2^{139}})$~\cite{Kyber-Kem} (as the failure probability for Kyber512 is $2^{-139}$).  
The decryption noise associated stays in between $(-\lfloor \frac{q}{8} \rfloor,\ \lfloor \frac{q}{8} \rfloor)$ with probability $(1-\frac{1}{2^{26}})$. Concretely it means that obtaining any noise value out of this range would happen only once in 67 million executions. From a physical attack perspective, it is a quite low probability.  
These observations indicate that (due to the bell curve of noise distribution) the noise values are highly likely to belong within a small range\footnote{\textcolor{black}{Note that all these analyses are for Kyber512 and the probability and ranges will slightly vary for other security versions of Kyber. However, the overall pattern remains the same thanks to the bell pattern of the distribution.}}.
} 

\textcolor{black}{Let us now analyze the distribution of the message coefficients ($g[i]$) just before the extraction of the message using the \texttt{A2B} algorithm (i.e. before line 4 in Algorithm~\ref{algo:masked-decode}). These coefficients contain the message added with the decryption noise.  
At the input of $\mathtt{A2B}$, $g[i] \in (0,\ \lfloor \frac{q}{2} \rfloor) = (0, 1664)$, if the corresponding message bit $m[i] = 0$, and $g[i] \in ( 2^{13} - \lfloor \frac{q}{2} \rfloor,\ 2^{13}-1) = (6528, 8191)$, when the corresponding message bit $m[i] = 1$. These two ranges are clearly disjoint between $1664$ and $6528$. For the other end (i.e. $0$ and $8191$), they are quite close as $0$ = $8192$ in the ring. The main observation at this point is that \emph{if a fault corrupts the message bit, then it must be strong enough to take the value of $g[i]$ from one range to the other range (e.g. from the value range of message value 0 to the value range of the message value 1).} As we can see, it is indeed possible if the value of $g[i]$ is close to $0$ or $8191$. However, such values occur with extremely low probability.}

\textcolor{black}{
\textcolor{black}{Now we consider the following scenarios, which model various single/multi-bit fault models on this variable $g[i]$. For the sake of explanation, we consider the fault width to be 8 bit, in this case, limited to the least significant bits (in fact, we mostly observe such faults in our practical setup). However, the argument is similar for any bit width. We note that the faults are actually injected on one of the shares of $g[i]$. However, the impact is realized on the unmasked $g[i]$ as the shares are added with \texttt{A2B}.} 
\begin{itemize}
    \item \textbf{\texttt{stuck-at-1} fault at least significant $8$ bits in a share of $g[i]$}: In this case, random bits in between $1$ to $8$ gets faulted to 1 if its value is 0. If the $j$-th bit is flipped from $0$ to $1$, then $2^{j-1}$ gets added with $g[i]$ after the fault. As we are introducing fault at a random share, the maximum added term to $g[i]$, that can occur from \texttt{stuck-at-1} fault at least significant $8$ bits in a share of $g[i]$, is $(1 + 2 + 4 + \cdots + 128) = 255$. \textcolor{black}{This happens when all of the 8 least significant bits in a share are 0, and all of them flip to 1. So the resulting value addition with $g[i]$ is 255.}   
    \item \textbf{\texttt{stuck-at-0} fault at least significant $8$ bits in a share of $g[i]$}:
    Here, random bits in between $1$ to $8$ get changed from $1$ to $0$. If the $j$-th bit is faulted from $1$ to $0$, then $2^{j-1}$ gets subtracted from $g[i]$ after the flip (as we are introducing fault at a masking share). So, the maximum subtracted term, that can occur from \texttt{stuck-at-0} fault at least significant $8$ bits in a share of $g[i]$, is $(1 + 2 + 4 + \cdots + 128) = 255$.
    \item \textbf{\texttt{bit-flip} fault at least significant $8$ bits  in a share of $g[i]$}: In this case, random bits in between $1$ to $8$ get flipped from either $0$ to $1$ or $1$ to $0$. If the $j$-th bit is flipped from $0$ to $1$, then $2^{j-1}$ gets added with $g[i]$ after the flip, and if the $j$-th bit is flipped from $1$ to $0$, then $2^{j-1}$ gets subtracted from $g[i]$ after the flip. So, the maximum added term that can occur in this case is $255$, as well as it is the maximum subtracted term that can occur.
\end{itemize}
}

\textcolor{black}{Let us first consider the scenario when the message bit $m[i]$ associated with $g[i]$ is 1. The decryption failure happens when we can change $g[i]$ to some values that decode to 1, and we want to do it with 8-bit fault models described before. This can indeed happen for $g[i] = 8191$ or values closer to it if we add $255$ with it. However, the least possible value for which this would be feasible is $7937$, which occurs with a very low probability. If we consider values from the higher probability range, such as $(6944,\ 7776)$ (happens with probability $1 - \frac{1}{2^{26}}$), the chances that it will go to the ranges where the message value is $0$ by adding (resp, subtracting) $255$ is zero. In other words, the probability of changing a message value from 1 to 0 with an 8-bit fault is definitely $ < \frac{1}{2^{26}}$, which we consider as quite low in terms of physical attacks. This is illustrated in Table.~\ref{tab:value-gi-after-fault}. In conclusion, we point out that with \emph{multi-bit faults at LSB positions up to 8 bits, it is highly improbable to change the value of the message bit. In other words, the faults at lower order bits never reach the 12/13th bit and, therefore, do not cause noise in our observations.}     
%
}

\textcolor{black}{It is worth mentioning that the probability of fault propagation to the message bit indeed increases if we consider multi-bit faults up to 9th bits. However, in simulation, we observed that the probability of such propagation is still low (e.g. no decryption failure for up to 9th-bit injection even if we simulate for 1 million cases). For faults up to 10/11 bits, the probability is significantly high for random bit-flip faults (nearly 50\%), but lower for stuck-at faults (nearly 30\%). The 11th bit is our signal and therefore, we always want a single bit fault individually occurring there. This also happens with some probability and in the experimental section, we show that it is indeed possible to increase this probability by careful adjustment of fault parameters. Finally, for $12$th and $13$th bits, the fault happens with $50\%$ probability and carries no information. The takeaway of this section, therefore, is we have to carefully avoid injecting faults in these MSB bits and keep the width of multi-bit faults low. Interestingly, most of the single-bit faults, even while occurring at random locations other than the 11/12/13th bit, never propagate. Overall, we see that the fault models can be significantly relaxed for our case, and even with a few single bit stuck-at faults hitting the 11th bit, we can perform the desired fault propagation. There are many situations where the probability of noise is low (or, in fact, zero), and if the fault model can be tuned for any one of those situations, the attack would work seamlessly.}

\begin{table}[!ht]
\centering
\caption{The effect of \texttt{stuck-at-1}, \texttt{stuck-at-0}, and \texttt{bit-flip} fault at the least significant $8$ bits of $g[i]$}
\label{tab:value-gi-after-fault}
\begin{tabular}{c|ccc|clc|c}
\hline
{\color[HTML]{000000} }                              & \multicolumn{3}{c|}{{\color[HTML]{000000} Before the fault injection}}                                                                                                                                                 & \multicolumn{3}{c|}{{\color[HTML]{000000} After the fault injection}}                                                                                                        & {\color[HTML]{000000} }                                                                                                                                  \\ \cline{2-7}
{\color[HTML]{000000} }                              & \multicolumn{1}{c|}{{\color[HTML]{000000} }}                                                                           & \multicolumn{2}{c|}{{\color[HTML]{000000} }}                                                  & \multicolumn{2}{c|}{{\color[HTML]{000000} }}                           & {\color[HTML]{000000} }                                                                             & {\color[HTML]{000000} }                                                                                                                                  \\
{\color[HTML]{000000} }                              & \multicolumn{1}{c|}{{\color[HTML]{000000} }}                                                                           & \multicolumn{2}{c|}{\multirow{-2}{*}{{\color[HTML]{000000} $g[i]$}}}                          & \multicolumn{2}{c|}{\multirow{-2}{*}{{\color[HTML]{000000} $g^*[i]$}}} & {\color[HTML]{000000} }                                                                             & {\color[HTML]{000000} }                                                                                                                                  \\ \cline{3-6}
\multirow{-4}{*}{{\color[HTML]{000000} Fault model}} & \multicolumn{1}{c|}{\multirow{-3}{*}{{\color[HTML]{000000} \begin{tabular}[c]{@{}c@{}}Message\\ $m[i]$\end{tabular}}}} & \multicolumn{1}{c|}{Min}                      & Max                                           & \multicolumn{1}{c|}{Min}          & \multicolumn{1}{c|}{Max}           & \multirow{-3}{*}{{\color[HTML]{000000} \begin{tabular}[c]{@{}c@{}}Message\\ $m^*[i]$\end{tabular}}} & \multirow{-4}{*}{{\color[HTML]{000000} \begin{tabular}[c]{@{}c@{}}Approximate \\ probability of the \\ fault-induced\\ decryption failure\end{tabular}}} \\ \hline
\multicolumn{1}{l|}{\texttt{stuck-at-1}}             & \multicolumn{1}{c|}{}                                                                                                  & {\color[HTML]{333333} }                       & {\color[HTML]{333333} }                       & \multicolumn{1}{l}{6944}          & \multicolumn{1}{l|}{8031}          & 1                                                                                                   &                                                                                                                                                          \\
\multicolumn{1}{l|}{\texttt{stuck-at-0}}             & \multicolumn{1}{c|}{}                                                                                                  & {\color[HTML]{333333} }                       & {\color[HTML]{333333} }                       & \multicolumn{1}{l}{6689}          & \multicolumn{1}{l|}{7776}          & 1                                                                                                   &                                                                                                                                                          \\
\multicolumn{1}{l|}{\texttt{bit-flip}}               & \multicolumn{1}{c|}{\multirow{-3}{*}{1}}                                                                               & \multirow{-3}{*}{{\color[HTML]{333333} 6944}} & \multirow{-3}{*}{{\color[HTML]{333333} 7776}} & \multicolumn{1}{l}{6689}          & \multicolumn{1}{l|}{8031}          & 1                                                                                                   & \multirow{-3}{*}{$\frac{1}{2^{26}}$}                                                                                                                     \\ \hline
\end{tabular}
\end{table}

%% file: sections/experiment.tex
\vspace{-15pt}
\section{Experiments}\label{sec:experiments}
\textcolor{black}{ In this section, we validate the attack on Kyber in simulation and also through practical fault injections. We begin with simulating the ideal scenario, i.e. injecting \texttt{stuck-at-1} at the $11$th bit of an arithmetic share of $g[i]$ for every execution of the decapsulation algorithm. The outcomes are then fed to the BP algorithm to recover the secret key. The simulated experiment helps us to find the right parameter settings for the BP algorithm. Next, we move to our practical setup with EM-based fault injections on a software implementation. We further tune the BP algorithm to perform key recovery from these practical fault cases.}     

\subsection{Noise Removal from Inequalities}\label{subsec:noise-removal}
\textcolor{black}{
Given a fault injected in the masked decapsulation algorithm with a ciphertext $C$, the decryption failure provides the information that $g[i]^{(11)} = 1$. The decryption success indicates that $g[i]^{(11)} = 0$.  \emph{The fault must be active in both cases}. However, in a single execution with $C$, it might happen that we observe successful decryption even though $g[i]^{(11)}=1$ due to the fact that the fault is not activated. Note that fault activation happens with probability $\frac{1}{2}$ since we fault the masked values. 
In order to remove the uncertainty (which results in wrong information and is interpreted as noise in the BP algorithm) due to probabilistic fault activation, our strategy is to execute the decapsulation of $C$ multiple times (of course, with fault injection). In our injection campaign, we repeat $C$, $\beta$ times until we see a decryption failure for it. In case no decryption failure is observed after $\beta$ executions, we decide $g[i]^{(11)}=0$. Once we see a decryption failure, we stop the execution with $C$, conclude $g[i]^{(11)}=1$, and move to the next ciphertext.}

\textcolor{black}{ In our simulation, we begin with $\beta = 20$. This is an empirically chosen value. However, we checked (through simulation) that the probability of getting a wrong decision about the value of $g[i]^{(11)}$ is negligible for this choice of $\beta$. Moreover, as mentioned \textcolor{black}{before}, the probability of $g[i]^{(11)}=1$ is significantly higher \textcolor{black}{($99.3\%$)} than $g[i]^{(11)}=0$, by the nature of the decryption noise distribution. A result of such bias in the distribution is that we get a very low number of decryption ``success'' cases compared to the decryption ``failure'' cases. Since we stop execution for a given $C$ upon seeing a decryption failure, the average number of repetitions we need is quite low compared to $\beta$.   
%
Empirically, we observed that, on average, we need to repeat our fault injection $2.67$ times for each ciphertext (We tried with $100$K ciphertext, and the total fault required there was $2.67$K). \textcolor{black}{The parameter setting for the ideal simulation is presented in Table~\ref{tab:parameter-set-inequality-solver}.}}

\begin{table}[http]
\centering
\scriptsize
\caption{The parameter settings of the BP algorithm for recovering the whole secret key}
\label{tab:parameter-set-inequality-solver}
\begin{tabular}{c|c|c|c|c}
\hline
\multirow{2}{*}{Fault model}                                                            & \multirow{2}{*}{Required inequalities} & \multirow{2}{*}{Different ciphertexts} & \multirow{2}{*}{\begin{tabular}[c]{@{}c@{}}Repetition \\ ($\beta$)\end{tabular}} & \multirow{2}{*}{Required Faults} \\
                                                                                        &                                        &                                        &                                                                                  &                                  \\ \hline
\multirow{2}{*}{\begin{tabular}[c]{@{}c@{}}Simulated ideal \\ fault model\end{tabular}} & \multirow{2}{*}{30,000}                & \multirow{2}{*}{60,000}                & \multirow{2}{*}{20}                                                              & \multirow{2}{*}{160,719}         \\
                                                                                        &                                        &                                        &                                                                                  &                                  \\ \hline
\multirow{2}{*}{\begin{tabular}[c]{@{}c@{}}Practical fault\\  model\end{tabular}}       & \multirow{2}{*}{35,000}                & \multirow{2}{*}{70,000}                & \multirow{2}{*}{180}                                                             & \multirow{2}{*}{1,857,294}       \\
                                                                                        &                                        &                                        &                                                                                  &                                  \\ \hline
\end{tabular}
\end{table}

\subsection{Configuring and Running the BP Algorithm}
\textcolor{black}{As mentioned in Section~\ref{sec:inequality-solve} we have utilized the inequality solver proposed by~\cite{Delvaux22}.}
\textcolor{black}{In our \texttt{stuck-at-1} fault injection attack at $g[i]^{(11)}$, we are only dealing with the inequalities that are generated from the message bit $m[i] = 1$. \textcolor{black}{This follows from the Lemmas in Section~\ref{subsec:fault-propagation}, which says that the fault propagates to the 13th bit only when the $12$th bit is 1. Since for Kyber, both the 12th and 13th bit are the same message bit, this means that the message bit must be 1. The corresponding inequalities for decryption success and failures are given in Equation~\ref{eq:ky2} and Equation~\ref{eq:ky1}, respectively, in Section~\ref{sec:inequality-solve}. However, due to such a highly unbalanced value distribution for $g[i]^{(11)}$, the BP solver might end up having a negligible number of inequalities corresponding to the correct cases compared to the faulty cases. The ratio is originally $0.7:99.3$, which improves to $3:97$ by applying a filtering strategy described in the next paragraph. This ratio, however, hinders the BP solving, making it converge to a wrong value quite often. To stabilize this, we introduce a strategy called \emph{sample rejection}. \emph{More precisely, we randomly reject $50\%$ of the inequalities among the total inequalities generated, but these rejected inequalities are only chosen from the decryption failure cases.} This improves the ratio between correct and faulty cases, making it $7: 93$. We found that this ratio is sufficient for consistently solving the system for several random secret key choices.
With this new balanced system of inequalities, we need roughly $30$k inequalities to recover the whole secret key (shown in Figure~\ref{fig:combined_graph}), and to obtain these $30$k inequalities (i.e. $60$k different ciphertexts, as we reject $50\%$ of the total inequalities and all are generated from decryption failures), we require to continue the fault simulation $160$k times (= repetitions required for noise removal $\times$ ciphertexts required to enable $50\%$ sample rejection = $2.67 \times 60$k). We present this result in Table~\ref{tab:parameter-set-inequality-solver}.
}   
}

\textcolor{black}{Another important parameter that controls the BP solving is the \emph{filtering} strategy originally presented in~\cite{Delvaux22}. Filtering basically chooses \textcolor{black}{the smallest value} among the message coefficients (i.e. $g[i]$) for fault injection in order to assist the BP algorithm. Also, in our case, it has some impact on improving the ratio of correct and faulty executions, as mentioned in the last paragraph. While directly implementing this strategy can be costly from the implementation perspective of the injection setup (as we need to change the fault location for every ciphertext), this is functionally similar to the following strategy: fixed the fault location at some coefficient index $i$, and then choose a ciphertext having the minimum value for this coefficient (among several randomly generated ciphertexts) for fault injection. This strategy, however, increases our simulation time, as we now need to generate several encapsulations to enable the minimum value selection. Owing to the similarity of the original filtering strategy, we, therefore, used the original filtering already implemented in the BP code. However, we take the minimum value for which the message bit is also 1. Also, it is worth mentioning that during the practical attack, we keep the fault location fixed, mimicking the second strategy.}

\subsection{Practical Setup}
\textcolor{black}{
After establishing the simulated evidence for the attack, we now move to the practical fault injection experiments. For our experiments, we target the \textcolor{black}{reference first-order masked} implementation of the considered scheme built for the ARM Cortex-M4 family of microcontrollers~\cite{FO-masked-kyber-HeinzKLPSS22}. We ported the reference implementation to the STM32F4-DISCOVERY board (DUT) housing the STM32F407 -- an ARM Cortex-M4 microcontroller running at a clock frequency of 24 MHz. \textcolor{black}{For the compilation of the reference implementation, we have
used arm-none-eabi-gcc version 10.3.1 with compiler flags \emph{-O3 -std=gnu99 -mthumb
-mcpu=cortex-m4 -mfloat-abi=hard -mfpu=fpv4-sp-d16}}. We have used a generic USB 2.0 to TTL converter for USART communication with the DUT. EM Fault Injection (EMFI) was utilized to inject faults into the target device.}

\begin{figure}[http]
    \centering
    \includegraphics[width=0.8\linewidth]{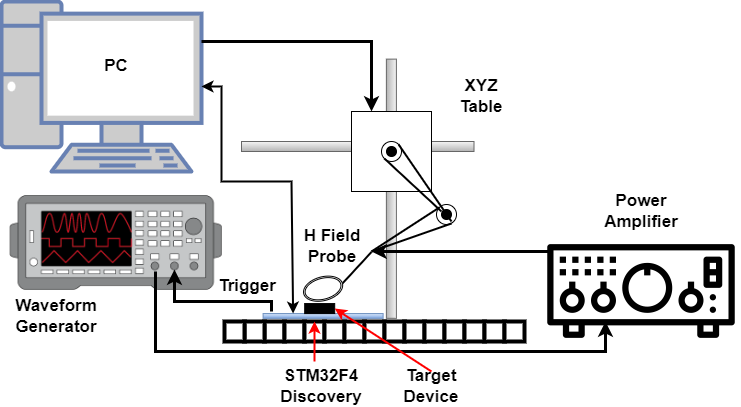}
    \caption{Schematic of the attack platform}
    \label{fig:fault_setup}
\end{figure}

\begin{figure}
    \centering
    \includegraphics[width=0.6\linewidth]{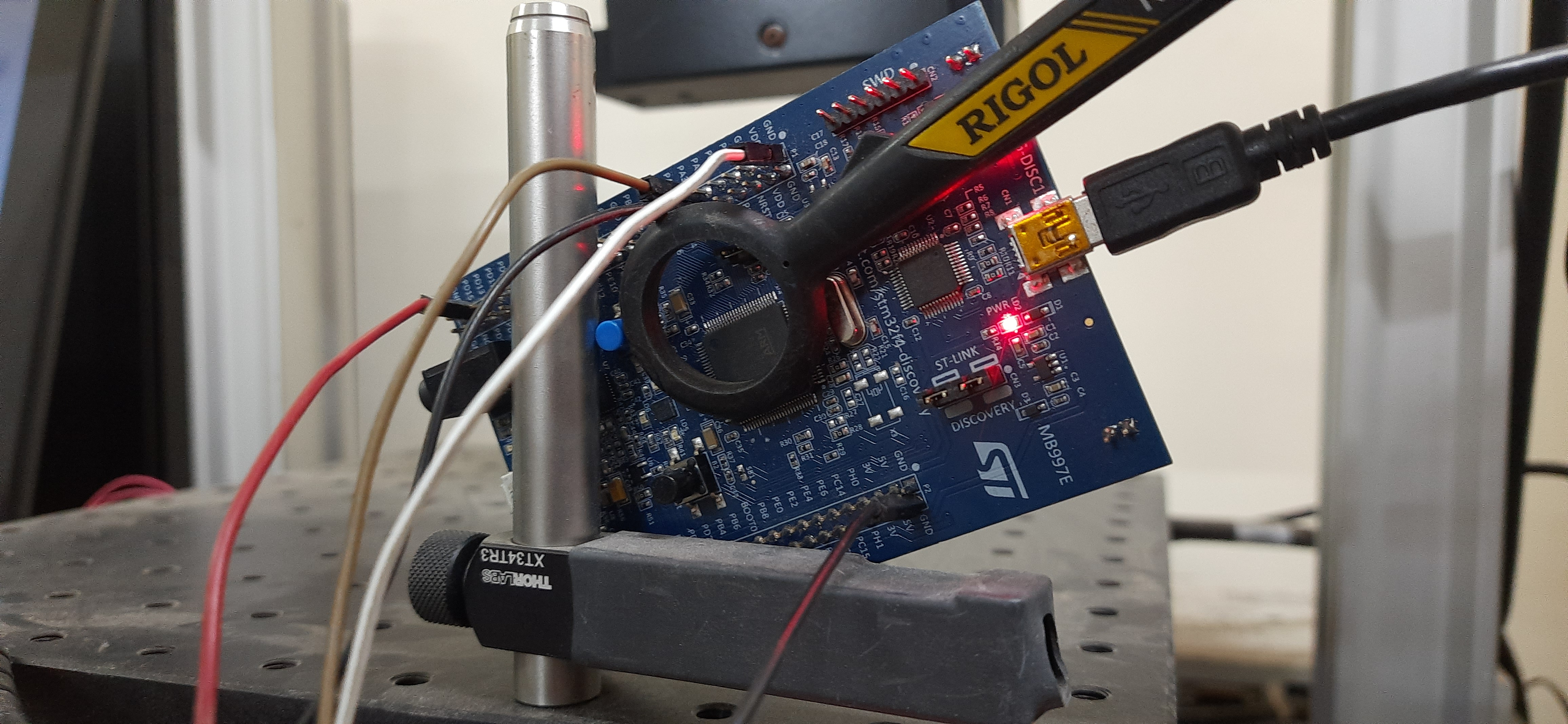}
    \caption{The position of the probe on the STM32F4 Discovery board}
    \label{fig:probe_pos}
\end{figure}

\textcolor{black}{The most formidable challenge in executing the aforementioned attack is inducing the faults accurately at desired locations. We have established an injection setup to address this challenge, illustrated in Figure \ref{fig:fault_setup}. This setup consists of several components, including an arbitrary waveform generator (Keysight 81160A), a constant-gain power amplifier (Teseq CBA 400M-260), a high-frequency near field H-probe (Rigol Near-field Probe 30MHz-3GHz), and an XYZ table (Thorlabs SMC100). Figure \ref{fig:probe_pos}
shows the position of the near-field EM probe on the DUT.} 

\textcolor{black}{
After being triggered by a signal from the evaluation board hosting the target implementation, the waveform generator produces a high-frequency pulse train. Following this, the amplifier boosts the intensity of this pulse train, and the H-probe generates a magnetic field above the target. It's important to highlight that the pulse train's amplitude, frequency, and burst count can be customized to meet particular needs. It was observed that by manipulating the position of the probe over the target using the XYZ table, with some necessary adjustments to the pulse parameters, fault can be induced in specific bits of the target register. However, this positioning process must rely on a trial-and-error approach due to the absence of internal register visibility. We have also utilized an oscilloscope to observe the execution timing of the target instruction so that the pulse delays can be adjusted accordingly and the trial-and-error can take the least amount of time. \textcolor{black}{Initially, we analyzed the target object file to get an insight into the occurrence of a particular instruction where we can inject the desired fault. In the \emph{masked\_poly\_tomsg}() function of the reference implementation, we have targeted a specific arithmetic instruction to induce the fault in the desired register, which corresponds to line 3 of Algorithm~\ref{algo:masked-decode}. } }



\subsection{Attack with Practical Faults}
\textcolor{black}{ \textcolor{black}{We do not expect the practically injected faults to exactly follow the ideal distribution (i.e. \texttt{stuck-at-1} fault at 11th bit with 100\% probability). It is more judicious to assume that the desired fault (\texttt{stuck-at-1} at $g[i]^{(11)}$) will occur with a certain probability $p$. If $p$ is significant, then we can expect some \emph{signal} (the desired fault event) to construct useful inequalities. Injections at undesired locations cause noise. However, we can handle such noise quite well, thanks to the two observations made in this paper: 1) A significant part of the induced faults on a share of $g[i]$ does not propagate to the $12$th or the $13$th bit (ref. Section~\ref{sec:relaxed-fault}), and 2) The uncertainty (noise) due to undesired injections can be largely removed through repeated execution of the same ciphertext (Section~\ref{subsec:noise-removal}). In Section~\ref{subsec:noise-removal}, the second observation was utilized to remove the uncertainty arising due to fault activation. However, the same strategy can be utilized to remove the uncertainties for undesired fault locations.}}

\begin{figure}[!ht]
    \centering
    \includegraphics[width=0.7\linewidth]{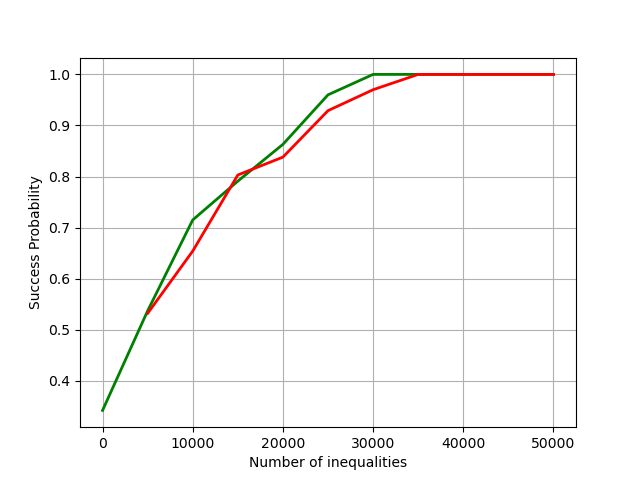}
    \caption{Number of inequalities required to recover partial or entire secret key for Kyber-512 with our simulated and practical fault model. Here, success probability indicates the number of coefficients recovered out of $1024 (= 512+512)$ coefficients of the secret key. (Success probability equal to $1$ means full secret key recovery.) The green coloured curve represents the simulated fault model, and the red coloured line represents the practical fault model. }
    \label{fig:combined_graph}
\end{figure}

\textcolor{black}{We, therefore, follow the same strategy as before: Upon repeating injection several ($\beta$) times for the same ciphertext $C$, if we see a decapsulation failure, we decide $g[i]^{11} = 1$. If we do not observe any decapsulation failure,  $g[i]^{11} = 0$. The choice of $\beta$ is crucial here, as a very low choice might lead to several wrong inequalities being added to the system. A very high choice, on the other hand, may increase the attack time. We note that even after choosing a proper $\beta$, we may end up having some wrong equations. However, the BP algorithm inherently tolerates some (low) noise. Therefore, it does not create an issue for us.}
\textcolor{black}{The most obvious (albeit slow) way or finding $\beta$ is to do some trial-and-error. We did this for the first set of outcomes from our practical setup with the same $\beta = 20$ value chosen for the ideal simulation. This did not work for obvious reasons. In order to speed up our experiments, we, therefore, extract the fault model directly from the setup, assuming an offline profiling stage. Analyzing faults at the target location, we observed that multi-bit faults were also induced, specifically towards the LSB side of the target register. After tweaking the pulse and burst parameters from the waveform generator, we were able to obtain $10\%$ single-bit stuck-at-1 faults at the 11\textsuperscript{th} bit and $90\%$ random single-bit/multi-bit stuck-at-1 faults, mostly concentrated at the bit positions $1-8$ (the LSB is at position 1).
\textcolor{black}{The pulse parameters are presented in Table~\ref{tab:pulse-parameters}.}
From this, we empirically estimate the value of $\beta$ as $180$. The interpretation is that if we repeat the fault injection $\beta$ times for the decapsulation of the same ciphertext, then the probability of the fault getting activated at the $11$th bit position at least once is significant. Interestingly, if the faults happen at any of the $1-8$ LSB positions, it does not propagate to the message bit ($13$th bit of $g[i]$)\footnote{\textcolor{black}{This is the reason why we mainly describe the 8-bit fault case in Section~\ref{sec:relaxed-fault}}.}. Therefore, with this observed fault model, we can ensure that if a decryption failure is observed, it is always due to the propagation of our desired fault. The $g[i]^{(11)}=0$ case, though, may have noise. However, $\beta = 180$ successfully removes this noise for most of the cases, enabling key recovery. We notice that, on average, for a single ciphertext, we need to repeat the fault injection $26.53$ times only. As shown in Figure~\ref{fig:combined_graph}, we could recover the secret key with $35$k inequalities even with this fault model. To obtain $35$k different balanced inequalities, we are required to continue the fault simulation $1,857$k times (= repetitions required for noise removal $\times$ ciphertexts required to enable $50\%$ sample rejection = $26.53 \times 70$k). We present this result in Table~\ref{tab:parameter-set-inequality-solver}. \textcolor{black}{It is interesting to observe that the number of inequalities is roughly the same for the ideal and the practical case. This is mainly attributed to the fault model we observe and the potentially removable noise by the structure of the attack.}}

\begin{table}[http]
\centering
\scriptsize
\caption{Pulse parameters}
\label{tab:pulse-parameters}
\begin{tabular}{|c|c|c|}
\hline
\multirow{2}{*}{Pulse Amplitude} & \multirow{2}{*}{Pulse Frequency} & \multirow{2}{*}{Burst Count} \\
                                 &                                  &                              \\ \hline
\multirow{2}{*}{-5.6 dBm}        & \multirow{2}{*}{200 MHz}         & \multirow{2}{*}{2}           \\
                                 &                                  &                              \\ \hline
\end{tabular}
\end{table}

%% file: sections/discussions.tex
\section{Discussion on Potential Countermeasures}\label{sec:discussion}
%
%
\textcolor{black}{
FA countermeasures use some form of redundant computation to detect/correct the fault \cite{Patranabis2018}. Considering the previous attacks on the LWE-based KEMs, many of them can be prevented using this strategy. For example, consider the attack due to~\cite{PesslP21}, which skips a subtraction operation. Duplication of the subtraction followed by a check can prevent such attacks. Quite similarly, the attack due to~\cite{PesslP21} can be prevented by duplicating different parts of the re-encryption module as the fault is mainly injected at that part. The reason why duplication works for these cases is that the attacks strongly rely on the \texttt{Decode} or the final check operation of the FO transform. Unless the fault reaches these operations, it cannot be exploited. Therefore, preventing the fault from reaching these operations can stop the attack.}

\textcolor{black}{In this regard, the proposed attack has some interesting properties. Although we still target the \texttt{Decode}, our injection happens much later than the fault location of the attack in~\cite{PesslP21}. In some sense, we target the penultimate step of the decoding operation itself rather than sending some faulty value to decode as in~\cite{HermelinkPP21,Delvaux22}. The most important fact is that if duplication and detection are used to detect our fault injection, that would result in the same information leakage that we utilize. Error correction might be somewhat useful, however, as shown in recent work~\cite{saha2021divided}, error correction too leaks if the FA is combined with SCA. We anticipate that recent combined-attack secure gadgets~\cite{FeldtkellerGMRSSS23} can be a proper direction to prevent this attack. However, such gadgets are based on Boolean or polynomial masking, while we target arithmetic masking. Shuffling can be useful in this context, but, again, the adversary can perform a combined attack to extract the permutation used for shuffling and thereby remove the noise induced by it. Note that, we already consider masking (even higher-order) for our target implementations. So, extracting only the shuffling permutation would not result in a successful SCA attack. But such extraction would definitely help our FA. With all these observations, we leave a sound countermeasure development as an interesting future work. }

%% file: sections/conclusion.tex
\vspace{-15pt}
\section{Conclusions}
\textcolor{black}{In this work, we propose a new FA on the SCA-secure masked decapsulation algorithm for generic LWE-based KEMs and elaborate the attack for Kyber. Our attack exploits the \texttt{A2B} component used in the masked implementations and appears due to the presence of masking. \textcolor{black}{So, it can be considered as a vulnerability introduced from the algorithmic changes introduced due to masking.} The attack results due to fault propagation through the \texttt{A2B} component and leaks a secret-dependent noise bit by unmasking it due to faults. Eventually, we show a practical validation of the attack on ARM Cortex-M4 microcontrollers using EMFI. A direct consequence of the attack is that one must be careful while designing masking algorithms. One also has to be careful while introducing fault countermeasures to prevent this attack, as we anticipate that state-of-the-art duplication-based countermeasures would not be sufficient.  
Overall, our work encourages more study in this direction, in general, to eventually construct truly secure implementations. 
%
}

\ifsubmission
\else
\section*{Acknowledgements}

This work was partially supported by Horizon 2020 ERC Advanced Grant (101020005 Belfort), CyberSecurity Research Flanders with reference number VR20192203, BE QCI: Belgian-QCI (3E230370) (see beqci.eu), and Intel Corporation. 
Angshuman Karmakar is funded by FWO (Research Foundation – Flanders) as a junior post-doctoral fellow (contract number 203056 / 1241722N LV). Debdeep would like to thank the ASEM DUO fellowship and the project entitled, "Secure Implementation of Post-Quantum Cryptosystems (SECPQC) TPN NO:- 71447", DST India and BELSPO for partial support. 

\fi